\documentclass[prb,aps,superscriptaddress,reprint,longbibliography]{revtex4-2}
\usepackage{siunitx}
\usepackage{graphicx}
\usepackage{amsmath}
\usepackage[english]{babel}
\usepackage[colorlinks=true,urlcolor=blue,linkcolor=blue,citecolor=blue]{hyperref}
\usepackage{mathptmx}
\usepackage{gensymb}

\usepackage{xcolor}
\usepackage[normalem]{ulem}
\usepackage{soul}

\begin{document}

\title{Chemically detaching hBN crystals grown at atmospheric pressure and high temperature \\ for high-performance graphene devices}

\author{Taoufiq Ouaj}
\email{taoufiq.ouaj@rwth-aachen.de}
\affiliation{JARA-FIT and 2nd Institute of Physics, RWTH Aachen University, 52074 Aachen, Germany}

\author{Leonard Kramme}
\affiliation{Institute of Inorganic Chemistry, RWTH Aachen University, 52056, Aachen, Germany}

\author{Marvin Metzelaars}
\affiliation{JARA-FIT and 2nd Institute of Physics, RWTH Aachen University, 52074 Aachen, Germany}
\affiliation{Institute of Inorganic Chemistry, RWTH Aachen University, 52056, Aachen, Germany}

\author{Jiahan Li}
\affiliation{Tim Taylor Department of Chemical Engineering, Kansas State University, Manhattan, Kansas 66506, United States}

\author{Kenji Watanabe}
\affiliation{Research Center for Functional Materials, National Institute for Materials Science, 1-1 Namiki, Tsukuba 305-0044, Japan}

\author{Takashi Taniguchi}
\affiliation{International Center for Materials Nanoarchitectonics, National Institute for Materials Science,  1-1 Namiki, Tsukuba 305-0044, Japan}

\author{James H. Edgar}
\affiliation{Tim Taylor Department of Chemical Engineering, Kansas State University, Manhattan, Kansas 66506, United States}

\author{Bernd Beschoten}
\email{bernd.beschoten@physik.rwth-aachen.de}
\affiliation{JARA-FIT and 2nd Institute of Physics, RWTH Aachen University, 52074 Aachen, Germany}
\affiliation{JARA-FIT Institute for Quantum Information, Forschungszentrum J\"ulich GmbH and RWTH Aachen University, 52074 Aachen, Germany}

\author{Paul K\"ogerler}
\affiliation{Institute of Inorganic Chemistry, RWTH Aachen University, 52056, Aachen, Germany}
\affiliation{Peter Gr\"unberg Institute, Electronic Properties (PGI-6) Forschungszentrum J\"ulich, 52425, J\"ulich, Germany}

\author{Christoph Stampfer}
\affiliation{JARA-FIT and 2nd Institute of Physics, RWTH Aachen University, 52074 Aachen, Germany}
\affiliation{Peter Gr\"unberg Institute (PGI-9) Forschungszentrum J\"ulich, 52425 J\"ulich, Germany}

\begin{abstract}

In this work, we report on the growth of hexagonal boron nitride (hBN) crystals from an iron flux at atmospheric pressure and high temperature and demonstrate that (i) the entire sheet of hBN crystals can be detached from the metal in a single step using hydrochloric acid and that (ii) these hBN crystals allow to fabricate high carrier mobility graphene-hBN devices. By combining spatially-resolved confocal Raman spectroscopy and electrical transport measurements, we confirm the excellent quality of these crystals for high-performance hBN-graphene-based van der Waals heterostructures. The full width at half maximum of the graphene Raman 2D peak is as low as $\mathrm{16\,cm^{-1}}$, and the room temperature charge carrier mobilitiy is around  $\mathrm{\num{80 000}\;\; cm^2/(Vs)}$ at a carrier density $\mathrm{\num{1e12}\,cm^{-12}}$. This is fully comparable with devices of similar dimensions fabricated using crystalline hBN synthesized by the high pressure and high temperature method. Finally, we show that for exfoliated high-quality hBN flakes with a thickness between 20 nm and 40 nm the line width of the 
hBN Raman peak, in contrast to the graphene 2D line width, is not useful for benchmarking hBN in high mobility graphene devices.

\end{abstract}

\maketitle

\section{Introduction}
Two-dimensional (2D) materials are found as metals, semi-metals, semiconductors, insulators or even magnets or superconductors~\cite{Gibertini2019May}, and they can be used to  create unparalleled
van der Waals (vdW) heterostructures, providing an exciting platform for both fundamental research and applications~\cite{Geim2013Jul, Novoselov2016Jul}.
Within the family of 2D materials, hexagonal boron nitride (hBN) excels due to its insulating properties (band gap of around ~5.9~eV~\cite{Watanabe2004Jun}), high thermal conductivity \cite{Lindsay2011Oct, Yuan2019May} and an ultra-flat surface that renders it an ideal substrate material for other 2D materials \cite{Dean2010Oct, Wang2013Nov}.
A large range of vdW heterostructures that rely on high-quality hBN have been developed, including graphene devices with ultra-high carrier mobilities~\cite{Dean2010Oct, Wang2013Nov, Banszerus2015Jul}, 
well-tunable bandgap systems in bilayer graphene~\cite{Icking2022Nov, Eich2018Aug} and high-performance optoelectronic devices based on transition metal dichalcogenides~\cite{Ajayi2017Jul, Raja2019Sep, Ersfeld2020May}.

The established method to obtain thin layers of crystalline hBN that meets the rigorous quality for high-performance devices involves the exfoliation (and dry transfer) from large, isolated (bulk) hBN crystals.~\cite{Meng2019Feb, Maestre2021Oct}.
Taniguchi and Watanabe pioneered the hBN crystal growth at high pressures ($\mathrm{>4\, GPa}$) and high temperatures (HPHT method) and provide the most significant source of high-quality hBN crystals within the 2D materials community~\cite{Taniguchi2007May, Zastrow2019Aug}.
However, crystals grown by this method may contain carbon-rich domains and their dimensions and production capability are limited by the design of the hydraulic press~\cite{Onodera2019Oct, Onodera2020Jan}.
The experimentally less demanding crystal growth at atmospheric pressure and high temperature (APHT method) from Ni-Mo and Ni-Cr ﬂuxes was first demonstrated by Kubota et al.~\cite{Kubota2007Aug, Kubota2008Mar, Kubota2007Jan}.
Subsequent research efforts led by Edgar's group~\cite{Hoffman2014May, Edgar2014Oct, Liu2018Sep,li2021Dec, Li2020} and later also by Wan's group~\cite{Zhang2019Nov, Cao2022Aug} have successfully optimized this method, resulting in the production of larger crystals of comparable quality.
Large crystalline hBN layers on top of the metal ingots could be achieved with single-crystalline domains ranging from  $\mathrm{100 \, \mu m}$ to $\mathrm{1 \,  mm}$ \cite{Maestre2021Oct, Li2020Jun, Li2021Apr}.
This demonstrates the potential of the APHT method to provide hBN crystals of high quality suitable for the fabrication of larger-scale devices based on graphene or other 2D materials~\cite{Naclerio2023Feb}.
However, removal of the crystal layers from the metal ingot via thermal release tape or mechanical force routinely fractures them into small ($\mathrm{mm^2}$-sized) flakes~\cite{Liu2018Sep, Liu2017Sep, Li2021Apr} and is limited in its "scalability".
Recently, Edgar and Liu both demonstrated that a single-component Fe flux is also suitable for the growth of large hBN single-crystal layers despite the predicted low solubility for nitrogen~\cite{Li2021Apr, Li2021Jul}.
Isolated hBN crystal flakes up to $\mathrm{4 \, cm^2}$ were recovered from the Fe ingot by exfoliating with thermal release tapes, while Li et al.~\cite{Li2021Jul} could demonstrate the removal of the entire as-grown crystal layer of similar size via mild chemical detachment by HCl treatment \cite{Li2021Jul}.
\begin{figure*}[t]
    \begin{center}
    \includegraphics{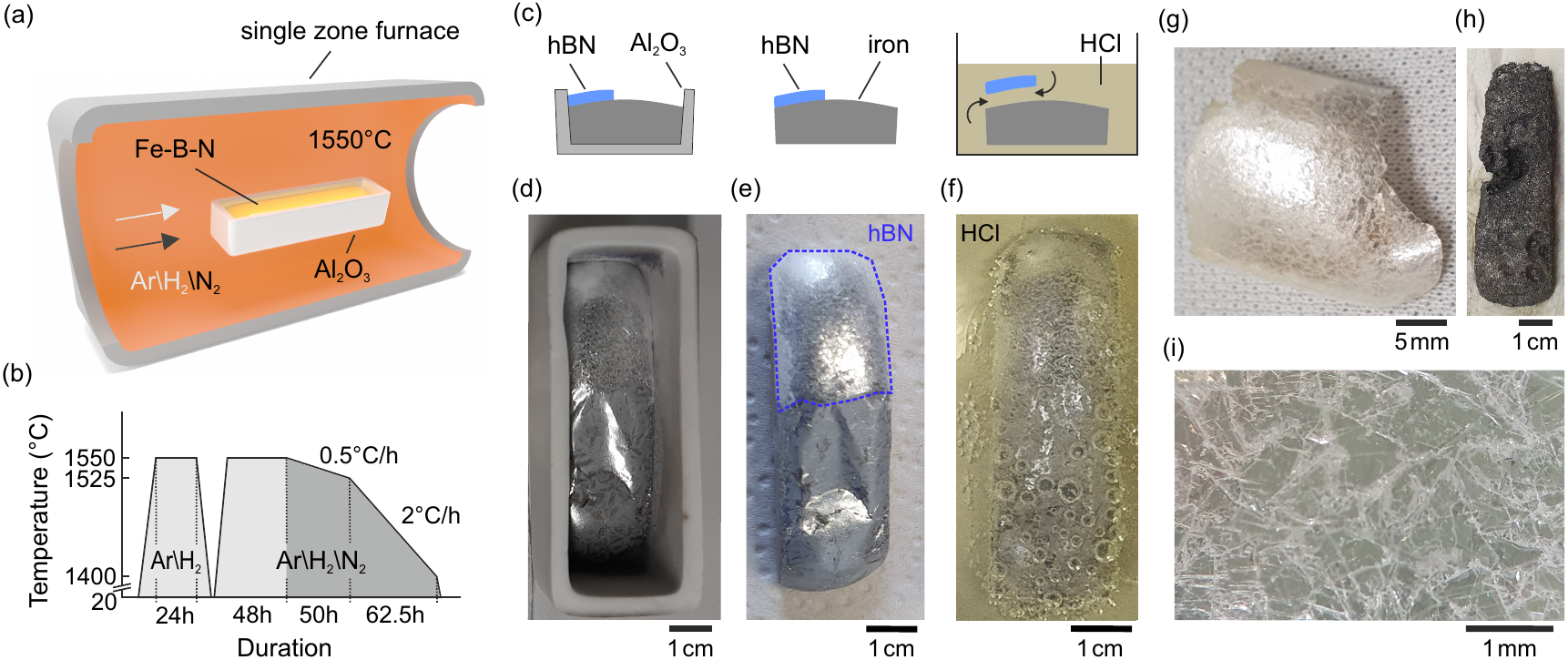}
    \end{center}
    \caption{Growth of hBN with the APHT method and detaching hBN crystals using hydrochloric acid. {\bf (a)} Schematic of the single-zone furnace at 1550°C with melted iron in an alumina crucible. {\bf(b)} Depiction of the temperature profile of the growth procedure: hBN crystals start to form after the saturation phase (indicated as a darker shade of grey).
    {\bf(c)} Schematic of the detaching process using hydrochloric acid. {\bf(d)-(f)} Optical images of the crucible with the iron ingot directly after growth (d), the iron ingot after taking it out of the crucible (e) and the iron ingot in the hydrochloric acid with emerging hydrogen bubbles (f). {\bf (g)} Optical image of the extracted hBN crystal and the remainder of the etched iron {\bf(h)}. {\bf (i)} Optical microscope image of a region on the crystal shown in (g).
  }
    \label{fig: extraction_method}
\end{figure*}
However, the integration of wet-chemistry in processing and handling 
2D materials are often susceptible to contaminations that may deteriorate the material quality \cite{Suk2011Sep}.
Therefore, it remains an open question whether an acid treatment that has also been used by Kubota \cite{Kubota2007Aug} and Li \cite{Li2021Jul} will affect the quality of the hBN crystals and more importantly their performance in state-of-the-art graphene-based vdW heterostructures.

Typical characterization methods to benchmark the quality of hBN crystals are based on confocal Raman spectroscopy as well as photo- and cathodoluminescence measurements~\cite{Schue2016Dec} that demonstrate similar or equal quality as crystals grown at high pressures.
A more direct approach to test the suitability of the hBN crystals for integration into high-performance graphene-based vdW heterostructures is the extraction of the electronic performance of an hBN-encapsulated graphene~\cite{Sonntag2020Jun, Onodera2019Oct, Neumann2015Sep}.

In this work, we present a holistic process of APHT hBN growth, high-yield detachment of the hBN crystal-layer from the solidified metal solution by HCl treatment, and subsequent integration of the hBN crystal-flakes for high-performance graphene-based vdW heterostructures fabricated by dry transfer. We confirm the high electronic performance of our devices by spatially-resolved confocal Raman microscopy and electronic transport measurements on structured Hall bar devices.
Temperature-dependent transport measurements reveal ballistic transport at low temperatures, while at room temperature, we identify "intrinsic" phonon scattering to limit the charge carrier mobility.
Moreover, we benchmark the suitability of hBN crystals grown by different methods (HPHT and APHT) under various conditions for their integration into graphene-based devices by comparing the hBN $\mathrm{E_{2g}}$  Raman peak width with the electronic properties of the hBN encapsulated graphene.

\section{Results and Discussion}
The hBN crystals that we used for the chemical detachment experiments and subsequent integration into graphene devices were grown from an Fe flux via the APHT method in a similar manner as described by Li et al.~\cite{Li2021Apr} and Li et al.~\cite{Li2021Jul}. The process is schematically depicted in Figs.~\ref{fig: extraction_method}(a,b).
Pieces of Fe (ChemPur, $\mathrm{0.5-1.5 \, mm}$, $\mathrm{99.97 \, \%}$) were mixed with $\mathrm{3.07\, wt\,\% }$ B powder (PI-KEM, $\mathrm{99.94 - 99.97\, \%}$) under inert conditions and loaded into a rectangular alumina crucible ($5\times2\times2$~cm$^{3}$) that was covered with a lid and placed into a horizontal tube furnace.
After three cycles of evacuation and refilling with $\mathrm{H_2}$ ($\mathrm{5 \, \%}$ in Ar) to $\mathrm{1.1 \, bar}$, the system was sintered at $\mathrm{1550 \, ^\circ C}$ for $\mathrm{24\, h}$ under a continuous gas flow of  $\mathrm{25\, sccm}$ to minimize oxygen and carbon impurities (see Fig.~\ref{fig: extraction_method}(b)).
The actual crystal growth process is performed under constant flows of $\mathrm{N_2}$ and $\mathrm{H_2}$ ($\mathrm{5 \, \%}$ in Ar) at $\mathrm{125\, sccm}$ and $\mathrm{25\, sccm}$, respectively, while maintaining a constant pressure of $\mathrm{1.1 \, bar}$.
After soaking at $\mathrm{1550 \, ^\circ C}$ for $\mathrm{48\, h}$ to saturate the metal flux with B and N, the furnace is slowly cooled to $\mathrm{1525 \, ^\circ C}$ with a rate of $\mathrm{0.5 \, ^\circ C/h}$ and subsequently to $\mathrm{1400 \, ^\circ C}$ with a rate of $\mathrm{2 \, ^\circ C/h}$.
Afterwards, the system is quickly cooled to room temperature at a rate of $\mathrm{300 \, ^\circ C/h}$.
A lower cooling rate should provide a lower density of nucleation sites and larger single-crystals but drastically increases the overall process time \cite{Hoffman2014May, Liu2018Sep}.
In our system, the combination of a slow crystallization phase ($\mathrm{0.5 \, ^\circ C/h}$) with a second, faster phase ($\mathrm{2 \, ^\circ C/h}$) produces more homogenous crystal layers while reducing the presence of small, detached precipitates.
A typical result of the growth process is schematically depicted in the left panel of Fig.~\ref{fig: extraction_method}(c) and in Fig.~\ref{fig: extraction_method}(d) we show the successful formation of a continuous hBN crystal layer covering half of the Fe ingot.
In accordance with the varied temperature program and boron content, our crystal layers appear to be thicker ($\mathrm{> 100 \, \mu m}$) than those reported by Edgar ($\mathrm{10-20 \mu m}$) and firmly adhere to the iron ingot.
This prevents a damage-free removal via thermal release tape and underlines the importance of the mild and reproducible detachment procedure that is schematically outlined in Fig.~\ref{fig: extraction_method}(c). 
After removing the iron ingot from the crucible (see Fig.~\ref{fig: extraction_method}(e)) and thoroughly investigating the attached crystalline hBN layer via optical microscopy and Raman spectroscopy (see below), it is immersed in concentrated HCl at room temperature.
The HCl slowly reacts with the iron ingot to form (solvated) $\mathrm{Fe^{2+}}$ and $\mathrm{Cl^-}$ ions and $\mathrm{H_2}$ gas as indicated by the bubble formation shown in Fig.~\ref{fig: extraction_method}(f).
After a reaction period that varies from batch to batch and may take several days, the HCl intercalates between the crystalline layer and the iron and facilitates a gentle detachment of the as-grown hBN crystal layer from the iron ingot that is aided by emerging $\mathrm{H_2}$ gas.
 The detached crystal layers were thoroughly cleaned with fresh HCl until no $\mathrm{FeCl_2}$ could be detected.
 After washing with distilled water and isopropanol the crystal was dried under nitrogen.
 The crystal layer is curved and has a dimension of $3$~cm$~\times$~3~cm (see Fig.~\ref{fig: extraction_method}(g)) with no visible damage or contamination from the detachment procedure and acid treatment.
 The remainder of the etched iron ingot is shown in Fig.~\ref{fig: extraction_method}(h).
 Microscopic analysis of the hBN crystal as shown in Fig.~\ref{fig: extraction_method}(i) reveals large crystal grains that are in the order of a few $\mathrm{100 \, \mu m}$.

\begin{figure}[t]
    \begin{center}
    \includegraphics[]{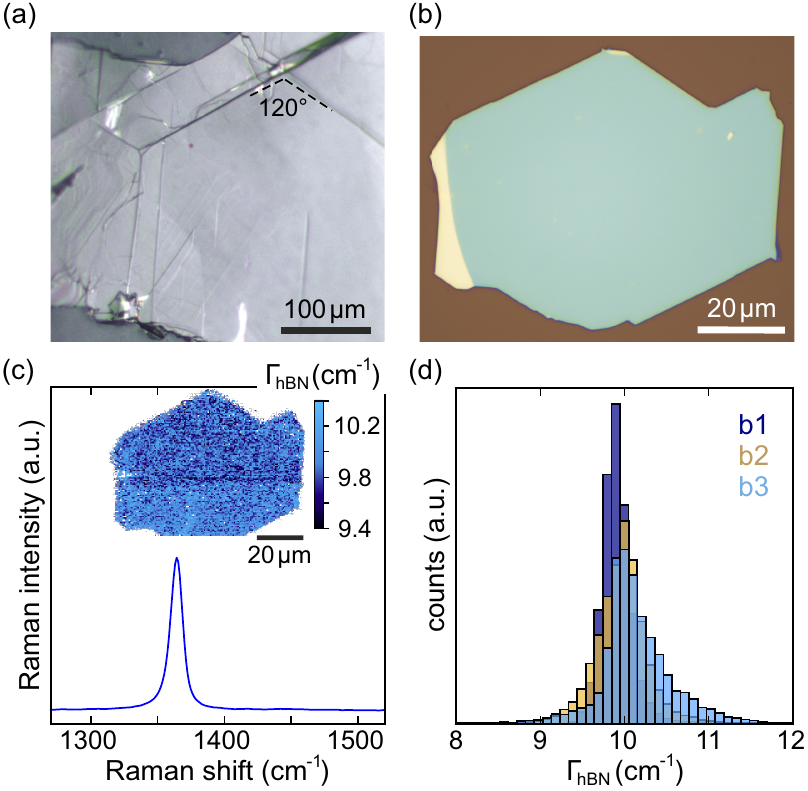}
    \end{center}
    \caption{Optical characterization of exfoliated hBN crystals. {\bf (a)} Optical microscope image of a hBN crystal on thermal release tape. {\bf(b)} Optical microscope image of an exfoliated flake on a silicon wafer with 90~nm SiO$_2$. {\bf(c)} Raman spectrum at one position of the flake shown in (b). The inset shows the spatially-resolved FWHM of the fitted $\mathrm{E_{2g}}$ Raman hBN peak, $\mathrm{\Gamma_{hBN}}$. {\bf(d)} Histograms showing data taken from spatially-resolved Raman maps taken on hBN crystals directly after detaching as well as on exfoliated flakes from these crystals. Three different batches (b1-b3), all grown using an iron flux and all detached using the same process, are compared, where all maps from the same batch were included in the respective histograms.
  }
    \label{fig: optical characterization}
\end{figure}

\begin{figure*}[t]
  \begin{center}
  \includegraphics{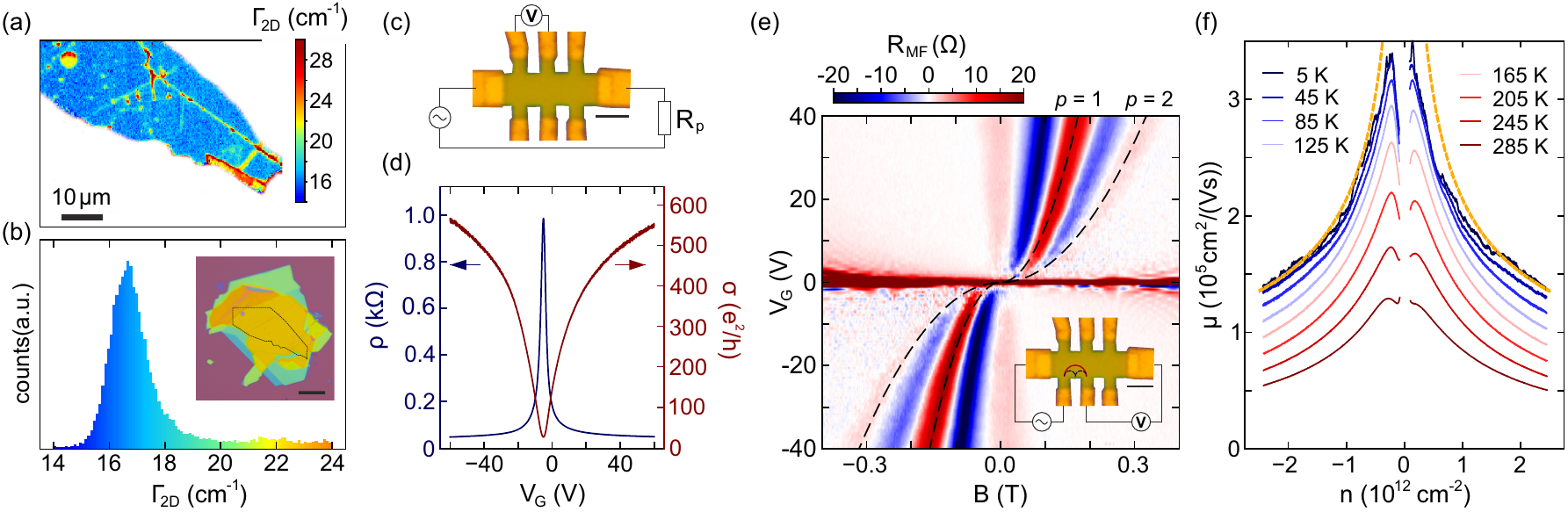}
  \end{center}
  \caption{Characterization of graphene encapsulated in hBN. {\bf (a)} Spatial distribution of the FWHM of the 2D Raman peak, $\mathrm{\Gamma_{2D}}$. {\bf (b)} Histogram of the FWHM of the Raman 2D peak shown in (a) and an optical image of the hBN-graphene-hBN heterostructure. The scale bar is 20~µm. {\bf (c)} Hall bar device with wiring scheme for the four-terminal measurements. The scale bar is 2~µm. {\bf (d)}~Four terminal conductivity and resistivity as function of the back gate voltage $V_\mathrm{G}$ (bias voltage of $V_\mathrm{b}=1$~V and a pre-resistance of $R\mathrm{_p = 1\, M\Omega}$ has been used).
  {\bf (e)}~Transverse magnetic focusing measurements taken at 1.7~K as function of magnetic field and gate voltage. The dashed curves correspond to the conditions where magnetic focusing works (see text). The inset shows the measurement configuration and the classical trajectories of the first two "modes". {\bf (f)}~Charge carrier mobility as function of carrier density for various temperatures. The yellow dashed line depicts the ballistic limit of the mobility, given roughly by the sample dimensions (see text). }
  \label{fig: graphene}
\end{figure*}

The quality of our hBN crystals was evaluated via spatially-resolved confocal Raman spectroscopy.
The evaluation involves mapping of the bulk hBN crystals directly after detachment and additionally after mechanical exfoliation of thin hBN flakes ($\mathrm{20-40\, nm}$) on silicon oxide.
To achieve this, a piece of hBN crystal is cleaved from the large crystal shown in Fig.~\ref{fig: extraction_method}(g) by pressing the top side of the crystal onto a tape repeatedly until a sufficient number of hBN flakes have been transferred to the tape.
The resulting hBN flakes on the tape have lateral sizes ranging from $\mathrm{100 \, \mu m}$ to $\mathrm{1 \, mm}$ (an example is shown in Fig.~\ref{fig: optical characterization}(a)), with a single crystal grain of hexagonal symmetry (see 120\degree$\;$ label in Fig.~\ref{fig: optical characterization}(a)) and side lengths of a few $\mathrm{100 \, \mu m}$.
To obtain thinner hBN flakes (below $\mathrm{40\, nm}$) suitable for integration into vdW heterostructures, the tape is folded multiple times and then pressed onto a silicon oxide wafer piece.
The wafer pieces are subsequently scanned using our automated flake search tool, where the thickness of the hBN flakes is determined via their optical contrast towards the substrate (see Fig.~5(f) in Ref.~\cite{Uslu2023Jun}).
An example of a suitable hBN flake is shown in Fig.~\ref{fig: optical characterization}(b), with a thickness of approximately $\mathrm{30\, nm}$ and a sufficiently large region of homogeneous thickness, making it a good candidate for graphene encapsulation.
In Fig.~\ref{fig: optical characterization}(c), a Raman spectrum of the exfoliated flake is presented, showing the $\mathrm{E_{2g}}$ hBN Raman peak at $\mathrm{1366\, cm^{-1}}$ with a full-width-at-half-maximum (FWHM) of $\mathrm{\Gamma_{hBN}} = \mathrm{9.8\, cm^{-1}}$.
The inset displays the spatial distribution of $\mathrm{\Gamma_{hBN}}$ for the hBN flake shown in Fig.~\ref{fig: optical characterization}(b), ranging from $\mathrm{9.5\, cm^{-1}}$ to $\mathrm{10.3\, cm^{-1}}$.

To gain a better understanding of the statistical distribution of the line width $\mathrm{\Gamma_{hBN}}$, we performed Raman measurements to map various regions of the crystals directly after exfoliation from three different growth batches using an iron melt, with minor variations in growth parameters.
The results are plotted in Fig.~\ref{fig: optical characterization}(d) with three histograms from the different batches (b1-b3) plotted on top of each other.
Each histogram represents data from at least five Raman maps of both exfoliated flakes or crystals directly after detachment.
The $\mathrm{\Gamma_{hBN}}$ values range from $\mathrm{9 \, cm^{-1}}$ to  $\mathrm{11 \, cm^{-1}}$, with mean values around $\mathrm{10 \, cm^{-1}}$, and are comparable to values obtained from hBN crystals prior to the acid treatment with HCl or to hBN flakes exfoliated using blue tape.
Values as low as $\mathrm{7.6 \, cm^{-1}}$ have been reported in literature~\cite{Cusco2020Apr} and we also observe similar values in crystals from other growth batches not shown here.
However, especially after exfoliation, the $\mathrm{\Gamma_{hBN}}$ values were around $\mathrm{10 \; cm^{-1}}$, which is also observed by other groups~\cite{Schue2016Dec}.
Our statistical evaluation, which yields only mean values below $\mathrm{10 \; cm^{-1}}$ for all investigated regions, therefore confirms the crystallinity across the entire detached crystal layer.
\\
Next, we use our hBN crystals to build encapsulated graphene devices and evaluated their electronic performance by examination of spatially-resolved Raman maps as well as electronic transport measurements at various temperatures between $\mathrm{5\, K}$ and $\mathrm{300\, K}$.
This allows us to use graphene to detect variations (inhomogeneities) in the quality of the hBN-graphene interface as well as to directly confirm the suitability of processed hBN crystals for high-performance graphene devices.
We employ standard dry transfer techniques based on a PDMS/PC stamp to transfer hBN and graphene flakes~\cite{Bisswanger2022Jun}.
Fig.~\ref{fig: graphene}(a) shows a spatially-resolved map of the FWHM of the graphene 2D Raman peak, $\mathrm{\Gamma_{2D}}$, of the heterostructure optically shown in the inset of Fig.~\ref{fig: graphene}(b).
We extract a $\mathrm{\Gamma_{2D}}$ line width down to $\mathrm{16 \; cm^{-1}}$ indicating minimal amount of nm-scale strain variations~\cite{Neumann2015Sep, Couto2014Oct} and low doping~\cite{Sonntag2023Feb}.
Areas of larger $\mathrm{\Gamma_{2D}}$ values are related to either bubbles, folds or contaminate accumulations between the layers, however, large areas of homogeneously low $\mathrm{\Gamma_{2D}}$ indicate high-quality of the involved hBN interfaces.

In the next step, Hall bars are structured in regions with no visible folds, bubbles, i.e. in regions with $\mathrm{\Gamma_{2D}} < \mathrm{17~ cm^{-1}}$.
We use electron beam lithography followed by evaporation of an aluminum mask and dry-etching with an $\mathrm{Ar/SF_6}$ atomic layer etching process.
The contacts are defined in a second lithography step with subsequent Cr/Au evaporation.
An example of a Hall bar structure and the four-terminal constant-current measurement scheme are shown in Fig.~\ref{fig: graphene}(c).
The highly doped Si substrate (covered by 285~nm SiO$_2$) is used as back gate (not shown).
Fig.~\ref{fig: graphene}(d) shows the room temperature (RT) gate-dependent four-terminal resistivity
$\rho=V_\mathrm{4T} W /(I\,L)$ (width $W = 2 \; \mu$m, length $L = 2 \;\mu$m and $I=1 \; \mu$A) and the conductivity $\sigma = 1/\rho$ measured on the sample shown in Fig.~\ref{fig: graphene}(c).
The conductivity exceeds $\mathrm{500 \; e^2/h}$ at large back gate voltages $V_\mathrm{G}$, i.e. at high charge carrier densities, mainly limited by electron-phonon scattering~\cite{Wang2013Nov}.
The charge neutrality point (CNP) is located at $V_\mathrm{G}^0 = -5.2 \;$V. Note that this trace was taken after multiple temperature cycles and that the initial position of the CNP was at $V_\mathrm{G}^0 = -0.4 \;$V. The shift of the CNP after multiple temperature cycles is also sometimes observed in other devices.

The high quality of our hBN-graphene heterostructure is further confirmed by conducting transverse magnetic focusing (MF) measurements at low temperature (1.7~K), as depicted in the inset of Fig.~\ref{fig: graphene}(e).
Charge carriers are injected between a smaller and larger contact and the response is measured non-locally as a function of gate voltage and magnetic field.
The Lorentz force bends the electrons in-plane, and when the cyclotron radius matches half of the distance between the contacts, the signal increases sharply.
The magnetic focusing response is expected to follow  $B_p = p \hbar \sqrt{n}/ (\sqrt{\pi} L)$, where $L$ is the distance between the injector and collector contact, $p-1$ is the number of reflections at the edge~\cite{Taychatanapat2013Apr} and  $n = \alpha (V_\mathrm{G}-V_\mathrm{G}^0)$, where $\alpha = 6.1 \times 10^{-12} \; \mathrm{cm^2/V}$ is the gate lever arm extracted from quantum Hall measurement (in agreement with the plate capacitor model) and $V_\mathrm{G}^0 = -80 \;$mV marks the charge neutrality point (at 1.7~K).
The dashed black curves in Fig.~\ref{fig: graphene}(e) show the expectations for $p = 1$ (no edge reflection) and $p = 2$ (one edge reflection), where $L=2\; \mu \mathrm{m}$ is the distance between the central points of the contacts.
The responses for $p = 1$ and $p = 2$ are strong and weak respectively, as expected for partly diffusive edge scattering~\cite{Taychatanapat2013Apr}, both in agreement with theory (see dashed lines).
The responses are equally pronounced for electrons $V_{\mathrm{G}}>0$ and holes $V_{\mathrm{G}}<0$, directly proving the ballistic nature of our device for both, electron and hole transport with a mean free path exceeding $\mathrm{3\; \mu m}$.
The large mean free path of the charge carriers is further confirmed by evaluating the temperature dependence of the mobility  $\mu = \sigma/(e n)$, which is depicted in Fig.~\ref{fig: graphene}(f), for temperatures between $\mathrm{5\, K}$ and $\mathrm{285\, K}$.
The yellow dashed line represents the mobility calculated using a mean free path of $\mathrm{2.5\; \mu m}$ which is on the same order of our sample size (i.e. sample width / length) and provides good indications that the ballistic limit is set by the device geometry.

\begin{figure}[t]
    \begin{center}
    \includegraphics{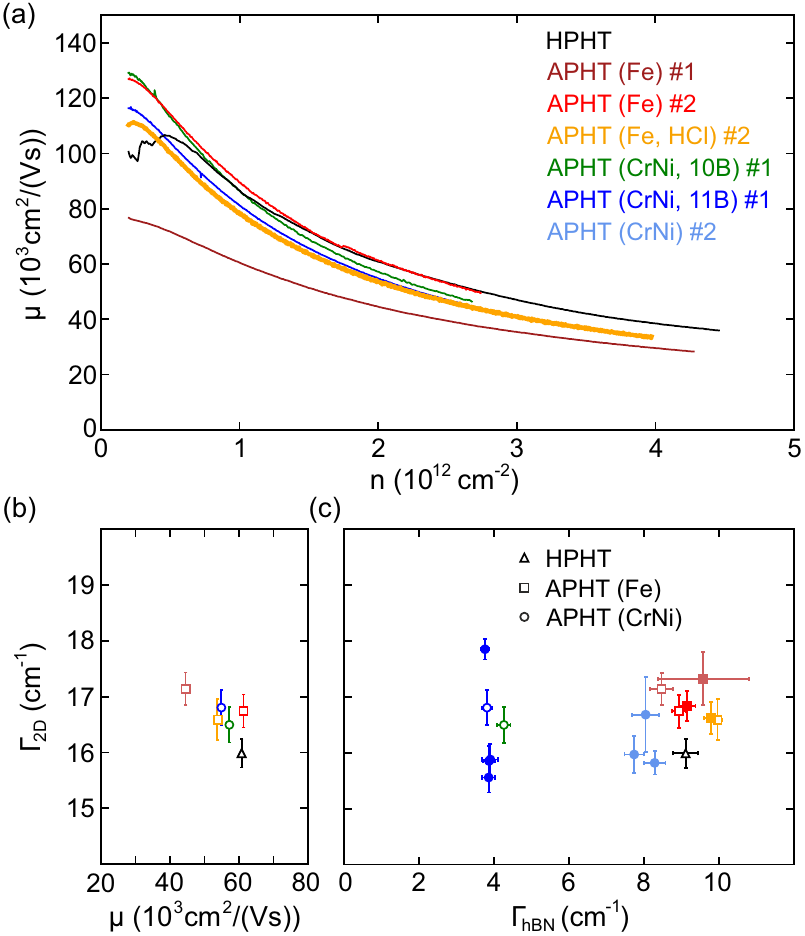}
    \end{center}
    \caption{Comparison of room temperature carrier mobilities. {\bf (a)}~Room temperature mobilities as function of charge carrier density for hBN-encapsulated graphene devices with hBN from different growth processes.
     {\bf (b)} The average FWHM of the graphene Raman 2D peak $\mathrm{\Gamma_{2D}}$ is plotted against the carrier mobility at $n=\mathrm{\num{2e12} \, cm^{-2}}$ (taken from (a)).  {\bf (c)}  Average $\mathrm{\Gamma_{2D}}$ as function of average the hBN peak $\mathrm{\Gamma_{hBN}}$. Values are extracted from homogeneous regions in the stack which were further used for Hall bar structuring. Open symbols correspond to the mobility traces in (a). }
    \label{fig: RT_mobility}
\end{figure}

Lastly, we compare the room temperature performance of a number of different hBN-encapsulated graphene samples using hBN grown in different batches and based on different growth parameters, metal fluxes, or the isotopic distribution of the boron.
The hBN crystals grown via the APHT method are either grown by the group of Edgar at KSU (\#1) or as described above at RWTH (\#2).
The Drude mobility for electron doping of several different devices is presented in Fig.~\ref{fig: RT_mobility}(a).
The device, based on HCl-detached hBN competes with other devices based on either HPHT hBN, $^\mathrm{10}$B or $^\mathrm{11}$B hBN as well as APHT hBN also grown from an iron flux (see labels in Fig.~\ref{fig: RT_mobility}(a)), but with a purely mechanical detachment of the hBN from the ingot.
Values of the Drude mobility up to $\mathrm{\num{120e3} \, cm^2(Vs)^{-1}}$ only limited by scattering with phonons are achieved.
However, we observe a sizable spread in the carrier mobility between devices from different growth batches, ranging from $\mathrm{\num{60e3} \, cm^2(Vs)^{-1}}$ to $\mathrm{\num{85e3} \, cm^2(Vs)^{-1}}$ at a charge carrier density of $n=\mathrm{\num{1e12} \, cm^{-2}}$
and around
$\mathrm{\num{40e3} \, cm^2(Vs)^{-1}}$ to $\mathrm{\num{65e3} \, cm^2(Vs)^{-1}}$ at $n = \mathrm{\num{2e12} \, cm^{-2}}$ making all these heterostructures interesting for high-performance electronic devices.
Interestingly, there is strong indication that the high carrier-density mobility of these devices is related just to the flatness of the graphene layer (i.e. the absence of nanometer-scale strain inhomogeneities), consistent with previous work on HPHT hBN \cite{Banszerus2015Jul,Couto2014Oct}.
This is shown in Fig.~\ref{fig: RT_mobility}(b), where the carrier mobility values appear to be (roughly) inversely proportional to the degree of strain inhomogeneities expressed by small average $\mathrm{\Gamma_{2D}}$ values, in agreement with Ref.~\cite{Couto2014Oct}.
It is noteworthy that the line width of the hBN Raman peak, $\mathrm{\Gamma_{hBN}}$, with average values up to $10 \; \mathrm {cm^{-1}}$ (see Fig.~\ref{fig: RT_mobility}(c)) is not linked to the carrier mobility in the nearby graphene layer.
Furthermore, the large difference in the hBN peak width between $^\mathrm{10}$B or $^\mathrm{11}$B and hBN with a natural isotopic distribution is related to the broadening due to isotopic disorder (mass effect)~\cite{Sonntag2020Jun}.
The absence of a relation between $\mathrm{\Gamma_{hBN}}$ and $\mathrm{\Gamma_{2D}}$ shows the limitation of using the hBN peak width as a measure of the quality of hBN crystals for benchmarking high-performance hBN-graphene heterostructure based devices.

\section{Conclusion and Outlook}

We conclude by stressing the importance of validating the quality of the surface and interfaces of exfoliated hBN crystals by directly probing the electronic quality of hBN-encapsulated graphene.
The use of graphene offers a sensitive way to benchmark various hBN crystals for their suitability as a substrate or encapsulant.
In addition, we have shown that high quality hBN crystals can be grown using the APHT method and successfully extracted with high yields and directly integrated into high performance graphene devices.
Most importantly, the electronic performance of graphene  encapsulated in HCl-etched hBN is as high as completely dry-processed hBN crystals.
Our work, therefore, suggests using certain wet-chemical process steps to increase the overall yield of high-quality hBN crystals is acceptable.
Interestingly, the result of the statistical analysis of spatially-resolved Raman maps has not let to any observable relation between the graphene 2D and hBN  peak widths for $\mathrm{\Gamma_{2D} < 18 \, cm^{-1}}$ and $\mathrm{\Gamma_{hBN} < 10 \, cm^{-1}}$. As a result, electronic transport measurements on hBN-encapsulated graphene devices remain the most important way to assess the quality of the hBN crystals involved.

\section{Data availability}
The data supporting the findings of this study are available in a Zenodo repository under https://doi.org/10.5281/zenodo.7799200, reference number 7799200.
\section{Acknowledgements}
The authors thank Luca Banszerus and Michael Schmitz for their helpful discussions and help with setting up the growth process at the beginning of this project, Jens Sonntag for his help with the electronic transport measurements and Jan-Lucas Uslu for helping in the fabrication process.
This project has received funding from the European Union’s Horizon 2020 research and innovation programme under grant agreement No. 881603 (Graphene Flagship) and from the European Research Council (ERC) under grant agreement No. 820254, the Deutsche Forschungsgemeinschaft (DFG, German Research Foundation) under Germany’s Excellence Strategy - Cluster of Excellence Matter and Light for Quantum Computing (ML4Q) EXC 2004/1 - 390534769, the FLAG-ERA grant TATTOOS, by the Deutsche Forschungsgemeinschaft (DFG, German Research Foundation) - 437214324  and by the Helmholtz Nano Facility~\cite{Albrecht2017}.
K.W. and T.T. acknowledge support from the JSPS KAKENHI (Grant Numbers 19H05790 and 20H00354).
In the USA, hBN crystal growth is supported by the Office of Naval Research, N00014-22-1-2582.
\section{Literature}


\begin{thebibliography}{45}%
\makeatletter
\providecommand \@ifxundefined [1]{%
 \@ifx{#1\undefined}
}%
\providecommand \@ifnum [1]{%
 \ifnum #1\expandafter \@firstoftwo
 \else \expandafter \@secondoftwo
 \fi
}%
\providecommand \@ifx [1]{%
 \ifx #1\expandafter \@firstoftwo
 \else \expandafter \@secondoftwo
 \fi
}%
\providecommand \natexlab [1]{#1}%
\providecommand \enquote  [1]{``#1''}%
\providecommand \bibnamefont  [1]{#1}%
\providecommand \bibfnamefont [1]{#1}%
\providecommand \citenamefont [1]{#1}%
\providecommand \href@noop [0]{\@secondoftwo}%
\providecommand \href [0]{\begingroup \@sanitize@url \@href}%
\providecommand \@href[1]{\@@startlink{#1}\@@href}%
\providecommand \@@href[1]{\endgroup#1\@@endlink}%
\providecommand \@sanitize@url [0]{\catcode `\\12\catcode `\$12\catcode
  `\&12\catcode `\#12\catcode `\^12\catcode `\_12\catcode `\%12\relax}%
\providecommand \@@startlink[1]{}%
\providecommand \@@endlink[0]{}%
\providecommand \url  [0]{\begingroup\@sanitize@url \@url }%
\providecommand \@url [1]{\endgroup\@href {#1}{\urlprefix }}%
\providecommand \urlprefix  [0]{URL }%
\providecommand \Eprint [0]{\href }%
\providecommand \doibase [0]{https://doi.org/}%
\providecommand \selectlanguage [0]{\@gobble}%
\providecommand \bibinfo  [0]{\@secondoftwo}%
\providecommand \bibfield  [0]{\@secondoftwo}%
\providecommand \translation [1]{[#1]}%
\providecommand \BibitemOpen [0]{}%
\providecommand \bibitemStop [0]{}%
\providecommand \bibitemNoStop [0]{.\EOS\space}%
\providecommand \EOS [0]{\spacefactor3000\relax}%
\providecommand \BibitemShut  [1]{\csname bibitem#1\endcsname}%
\let\auto@bib@innerbib\@empty
\bibitem [{\citenamefont {Gibertini}\ \emph {et~al.}(2019)\citenamefont
  {Gibertini}, \citenamefont {Koperski}, \citenamefont {Morpurgo},\ and\
  \citenamefont {Novoselov}}]{Gibertini2019May}%
  \BibitemOpen
  \bibfield  {author} {\bibinfo {author} {\bibfnamefont {M.}~\bibnamefont
  {Gibertini}}, \bibinfo {author} {\bibfnamefont {M.}~\bibnamefont {Koperski}},
  \bibinfo {author} {\bibfnamefont {A.~F.}\ \bibnamefont {Morpurgo}},\ and\
  \bibinfo {author} {\bibfnamefont {K.~S.}\ \bibnamefont {Novoselov}},\
  }\bibfield  {title} {\bibinfo {title} {{Magnetic 2D materials and
  heterostructures}},\ }\href {https://doi.org/10.1038/s41565-019-0438-6}
  {\bibfield  {journal} {\bibinfo  {journal} {Nat. Nanotechnol.}\ }\textbf
  {\bibinfo {volume} {14}},\ \bibinfo {pages} {408} (\bibinfo {year}
  {2019})}\BibitemShut {NoStop}%
\bibitem [{\citenamefont {Geim}\ and\ \citenamefont
  {Grigorieva}(2013)}]{Geim2013Jul}%
  \BibitemOpen
  \bibfield  {author} {\bibinfo {author} {\bibfnamefont {A.~K.}\ \bibnamefont
  {Geim}}\ and\ \bibinfo {author} {\bibfnamefont {I.~V.}\ \bibnamefont
  {Grigorieva}},\ }\bibfield  {title} {\bibinfo {title} {{Van der Waals
  heterostructures}},\ }\href {https://doi.org/10.1038/nature12385} {\bibfield
  {journal} {\bibinfo  {journal} {Nature}\ }\textbf {\bibinfo {volume} {499}},\
  \bibinfo {pages} {419} (\bibinfo {year} {2013})}\BibitemShut {NoStop}%
\bibitem [{\citenamefont {Novoselov}\ \emph {et~al.}(2016)\citenamefont
  {Novoselov}, \citenamefont {Mishchenko}, \citenamefont {Carvalho},\ and\
  \citenamefont {Neto}}]{Novoselov2016Jul}%
  \BibitemOpen
  \bibfield  {author} {\bibinfo {author} {\bibfnamefont {K.~S.}\ \bibnamefont
  {Novoselov}}, \bibinfo {author} {\bibfnamefont {A.}~\bibnamefont
  {Mishchenko}}, \bibinfo {author} {\bibfnamefont {A.}~\bibnamefont
  {Carvalho}},\ and\ \bibinfo {author} {\bibfnamefont {A.~H.~C.}\ \bibnamefont
  {Neto}},\ }\bibfield  {title} {\bibinfo {title} {{2D materials and van der
  Waals heterostructures}},\ }\href {https://doi.org/10.1126/science.aac9439}
  {\bibfield  {journal} {\bibinfo  {journal} {Science}\ }\textbf {\bibinfo
  {volume} {353}},\ \bibinfo {pages} {aac9439} (\bibinfo {year}
  {2016})}\BibitemShut {NoStop}%
\bibitem [{\citenamefont {Watanabe}\ \emph {et~al.}(2004)\citenamefont
  {Watanabe}, \citenamefont {Taniguchi},\ and\ \citenamefont
  {Kanda}}]{Watanabe2004Jun}%
  \BibitemOpen
  \bibfield  {author} {\bibinfo {author} {\bibfnamefont {K.}~\bibnamefont
  {Watanabe}}, \bibinfo {author} {\bibfnamefont {T.}~\bibnamefont
  {Taniguchi}},\ and\ \bibinfo {author} {\bibfnamefont {H.}~\bibnamefont
  {Kanda}},\ }\bibfield  {title} {\bibinfo {title} {{Direct-bandgap properties
  and evidence for ultraviolet lasing of hexagonal boron nitride single
  crystal}},\ }\href {https://doi.org/10.1038/nmat1134} {\bibfield  {journal}
  {\bibinfo  {journal} {Nat. Mater.}\ }\textbf {\bibinfo {volume} {3}},\
  \bibinfo {pages} {404} (\bibinfo {year} {2004})}\BibitemShut {NoStop}%
\bibitem [{\citenamefont {Lindsay}\ and\ \citenamefont
  {Broido}(2011)}]{Lindsay2011Oct}%
  \BibitemOpen
  \bibfield  {author} {\bibinfo {author} {\bibfnamefont {L.}~\bibnamefont
  {Lindsay}}\ and\ \bibinfo {author} {\bibfnamefont {D.~A.}\ \bibnamefont
  {Broido}},\ }\bibfield  {title} {\bibinfo {title} {{Enhanced thermal
  conductivity and isotope effect in single-layer hexagonal boron nitride}},\
  }\href {https://doi.org/10.1103/PhysRevB.84.155421} {\bibfield  {journal}
  {\bibinfo  {journal} {Phys. Rev. B}\ }\textbf {\bibinfo {volume} {84}},\
  \bibinfo {pages} {155421} (\bibinfo {year} {2011})}\BibitemShut {NoStop}%
\bibitem [{\citenamefont {Yuan}\ \emph {et~al.}(2019)\citenamefont {Yuan},
  \citenamefont {Li}, \citenamefont {Lindsay}, \citenamefont {Cherns},
  \citenamefont {Pomeroy}, \citenamefont {Liu}, \citenamefont {Edgar},\ and\
  \citenamefont {Kuball}}]{Yuan2019May}%
  \BibitemOpen
  \bibfield  {author} {\bibinfo {author} {\bibfnamefont {C.}~\bibnamefont
  {Yuan}}, \bibinfo {author} {\bibfnamefont {J.}~\bibnamefont {Li}}, \bibinfo
  {author} {\bibfnamefont {L.}~\bibnamefont {Lindsay}}, \bibinfo {author}
  {\bibfnamefont {D.}~\bibnamefont {Cherns}}, \bibinfo {author} {\bibfnamefont
  {J.~W.}\ \bibnamefont {Pomeroy}}, \bibinfo {author} {\bibfnamefont
  {S.}~\bibnamefont {Liu}}, \bibinfo {author} {\bibfnamefont {J.~H.}\
  \bibnamefont {Edgar}},\ and\ \bibinfo {author} {\bibfnamefont
  {M.}~\bibnamefont {Kuball}},\ }\bibfield  {title} {\bibinfo {title}
  {{Modulating the thermal conductivity in hexagonal boron nitride via
  controlled boron isotope concentration}},\ }\href
  {https://doi.org/10.1038/s42005-019-0145-5} {\bibfield  {journal} {\bibinfo
  {journal} {Commun. Phys.}\ }\textbf {\bibinfo {volume} {2}},\ \bibinfo
  {pages} {43} (\bibinfo {year} {2019})}\BibitemShut {NoStop}%
\bibitem [{\citenamefont {Dean}\ \emph {et~al.}(2010)\citenamefont {Dean},
  \citenamefont {Young}, \citenamefont {Meric}, \citenamefont {Lee},
  \citenamefont {Wang}, \citenamefont {Sorgenfrei}, \citenamefont {Watanabe},
  \citenamefont {Taniguchi}, \citenamefont {Kim}, \citenamefont {Shepard},\
  and\ \citenamefont {Hone}}]{Dean2010Oct}%
  \BibitemOpen
  \bibfield  {author} {\bibinfo {author} {\bibfnamefont {C.~R.}\ \bibnamefont
  {Dean}}, \bibinfo {author} {\bibfnamefont {A.~F.}\ \bibnamefont {Young}},
  \bibinfo {author} {\bibfnamefont {I.}~\bibnamefont {Meric}}, \bibinfo
  {author} {\bibfnamefont {C.}~\bibnamefont {Lee}}, \bibinfo {author}
  {\bibfnamefont {L.}~\bibnamefont {Wang}}, \bibinfo {author} {\bibfnamefont
  {S.}~\bibnamefont {Sorgenfrei}}, \bibinfo {author} {\bibfnamefont
  {K.}~\bibnamefont {Watanabe}}, \bibinfo {author} {\bibfnamefont
  {T.}~\bibnamefont {Taniguchi}}, \bibinfo {author} {\bibfnamefont
  {P.}~\bibnamefont {Kim}}, \bibinfo {author} {\bibfnamefont {K.~L.}\
  \bibnamefont {Shepard}},\ and\ \bibinfo {author} {\bibfnamefont
  {J.}~\bibnamefont {Hone}},\ }\bibfield  {title} {\bibinfo {title} {{Boron
  nitride substrates for high-quality graphene electronics}},\ }\href
  {https://doi.org/10.1038/nnano.2010.172} {\bibfield  {journal} {\bibinfo
  {journal} {Nat. Nanotechnol.}\ }\textbf {\bibinfo {volume} {5}},\ \bibinfo
  {pages} {722} (\bibinfo {year} {2010})}\BibitemShut {NoStop}%
\bibitem [{\citenamefont {Wang}\ \emph {et~al.}(2013)\citenamefont {Wang},
  \citenamefont {Meric}, \citenamefont {Huang}, \citenamefont {Gao},
  \citenamefont {Gao}, \citenamefont {Tran}, \citenamefont {Taniguchi},
  \citenamefont {Watanabe}, \citenamefont {Campos}, \citenamefont {Muller},
  \citenamefont {Guo}, \citenamefont {Kim}, \citenamefont {Hone}, \citenamefont
  {Shepard},\ and\ \citenamefont {Dean}}]{Wang2013Nov}%
  \BibitemOpen
  \bibfield  {author} {\bibinfo {author} {\bibfnamefont {L.}~\bibnamefont
  {Wang}}, \bibinfo {author} {\bibfnamefont {I.}~\bibnamefont {Meric}},
  \bibinfo {author} {\bibfnamefont {P.~Y.}\ \bibnamefont {Huang}}, \bibinfo
  {author} {\bibfnamefont {Q.}~\bibnamefont {Gao}}, \bibinfo {author}
  {\bibfnamefont {Y.}~\bibnamefont {Gao}}, \bibinfo {author} {\bibfnamefont
  {H.}~\bibnamefont {Tran}}, \bibinfo {author} {\bibfnamefont {T.}~\bibnamefont
  {Taniguchi}}, \bibinfo {author} {\bibfnamefont {K.}~\bibnamefont {Watanabe}},
  \bibinfo {author} {\bibfnamefont {L.~M.}\ \bibnamefont {Campos}}, \bibinfo
  {author} {\bibfnamefont {D.~A.}\ \bibnamefont {Muller}}, \bibinfo {author}
  {\bibfnamefont {J.}~\bibnamefont {Guo}}, \bibinfo {author} {\bibfnamefont
  {P.}~\bibnamefont {Kim}}, \bibinfo {author} {\bibfnamefont {J.}~\bibnamefont
  {Hone}}, \bibinfo {author} {\bibfnamefont {K.~L.}\ \bibnamefont {Shepard}},\
  and\ \bibinfo {author} {\bibfnamefont {C.~R.}\ \bibnamefont {Dean}},\
  }\bibfield  {title} {\bibinfo {title} {{One-Dimensional Electrical Contact to
  a Two-Dimensional Material}},\ }\href
  {https://doi.org/10.1126/science.1244358} {\bibfield  {journal} {\bibinfo
  {journal} {Science}\ }\textbf {\bibinfo {volume} {342}},\ \bibinfo {pages}
  {614} (\bibinfo {year} {2013})}\BibitemShut {NoStop}%
\bibitem [{\citenamefont {Banszerus}\ \emph {et~al.}(2015)\citenamefont
  {Banszerus}, \citenamefont {Schmitz}, \citenamefont {Engels}, \citenamefont
  {Dauber}, \citenamefont {Oellers}, \citenamefont {Haupt}, \citenamefont
  {Watanabe}, \citenamefont {Taniguchi}, \citenamefont {Beschoten},\ and\
  \citenamefont {Stampfer}}]{Banszerus2015Jul}%
  \BibitemOpen
  \bibfield  {author} {\bibinfo {author} {\bibfnamefont {L.}~\bibnamefont
  {Banszerus}}, \bibinfo {author} {\bibfnamefont {M.}~\bibnamefont {Schmitz}},
  \bibinfo {author} {\bibfnamefont {S.}~\bibnamefont {Engels}}, \bibinfo
  {author} {\bibfnamefont {J.}~\bibnamefont {Dauber}}, \bibinfo {author}
  {\bibfnamefont {M.}~\bibnamefont {Oellers}}, \bibinfo {author} {\bibfnamefont
  {F.}~\bibnamefont {Haupt}}, \bibinfo {author} {\bibfnamefont
  {K.}~\bibnamefont {Watanabe}}, \bibinfo {author} {\bibfnamefont
  {T.}~\bibnamefont {Taniguchi}}, \bibinfo {author} {\bibfnamefont
  {B.}~\bibnamefont {Beschoten}},\ and\ \bibinfo {author} {\bibfnamefont
  {C.}~\bibnamefont {Stampfer}},\ }\bibfield  {title} {\bibinfo {title}
  {{Ultrahigh-mobility graphene devices from chemical vapor deposition on
  reusable copper}},\ }\href {https://doi.org/10.1126/sciadv.1500222}
  {\bibfield  {journal} {\bibinfo  {journal} {Sci. Adv.}\ }\textbf {\bibinfo
  {volume} {1}},\ \bibinfo {pages} {e1500222} (\bibinfo {year}
  {2015})}\BibitemShut {NoStop}%
\bibitem [{\citenamefont {Icking}\ \emph {et~al.}(2022)\citenamefont {Icking},
  \citenamefont {Banszerus}, \citenamefont
  {W{\ifmmode\ddot{o}\else\"{o}\fi}rtche}, \citenamefont {Volmer},
  \citenamefont {Schmidt}, \citenamefont {Steiner}, \citenamefont {Engels},
  \citenamefont {Hesselmann}, \citenamefont {Goldsche}, \citenamefont
  {Watanabe}, \citenamefont {Taniguchi}, \citenamefont {Volk}, \citenamefont
  {Beschoten},\ and\ \citenamefont {Stampfer}}]{Icking2022Nov}%
  \BibitemOpen
  \bibfield  {author} {\bibinfo {author} {\bibfnamefont {E.}~\bibnamefont
  {Icking}}, \bibinfo {author} {\bibfnamefont {L.}~\bibnamefont {Banszerus}},
  \bibinfo {author} {\bibfnamefont {F.}~\bibnamefont
  {W{\ifmmode\ddot{o}\else\"{o}\fi}rtche}}, \bibinfo {author} {\bibfnamefont
  {F.}~\bibnamefont {Volmer}}, \bibinfo {author} {\bibfnamefont
  {P.}~\bibnamefont {Schmidt}}, \bibinfo {author} {\bibfnamefont
  {C.}~\bibnamefont {Steiner}}, \bibinfo {author} {\bibfnamefont
  {S.}~\bibnamefont {Engels}}, \bibinfo {author} {\bibfnamefont
  {J.}~\bibnamefont {Hesselmann}}, \bibinfo {author} {\bibfnamefont
  {M.}~\bibnamefont {Goldsche}}, \bibinfo {author} {\bibfnamefont
  {K.}~\bibnamefont {Watanabe}}, \bibinfo {author} {\bibfnamefont
  {T.}~\bibnamefont {Taniguchi}}, \bibinfo {author} {\bibfnamefont
  {C.}~\bibnamefont {Volk}}, \bibinfo {author} {\bibfnamefont {B.}~\bibnamefont
  {Beschoten}},\ and\ \bibinfo {author} {\bibfnamefont {C.}~\bibnamefont
  {Stampfer}},\ }\bibfield  {title} {\bibinfo {title} {{Transport Spectroscopy
  of Ultraclean Tunable Band Gaps in Bilayer Graphene}},\ }\href
  {https://doi.org/10.1002/aelm.202200510} {\bibfield  {journal} {\bibinfo
  {journal} {Adv. Electron. Mater.}\ }\textbf {\bibinfo {volume} {8}},\
  \bibinfo {pages} {2200510} (\bibinfo {year} {2022})}\BibitemShut {NoStop}%
\bibitem [{\citenamefont {Eich}\ \emph {et~al.}(2018)\citenamefont {Eich},
  \citenamefont {Pisoni}, \citenamefont {Pally}, \citenamefont {Overweg},
  \citenamefont {Kurzmann}, \citenamefont {Lee}, \citenamefont {Rickhaus},
  \citenamefont {Watanabe}, \citenamefont {Taniguchi}, \citenamefont
  {Ensslin},\ and\ \citenamefont {Ihn}}]{Eich2018Aug}%
  \BibitemOpen
  \bibfield  {author} {\bibinfo {author} {\bibfnamefont {M.}~\bibnamefont
  {Eich}}, \bibinfo {author} {\bibfnamefont {R.}~\bibnamefont {Pisoni}},
  \bibinfo {author} {\bibfnamefont {A.}~\bibnamefont {Pally}}, \bibinfo
  {author} {\bibfnamefont {H.}~\bibnamefont {Overweg}}, \bibinfo {author}
  {\bibfnamefont {A.}~\bibnamefont {Kurzmann}}, \bibinfo {author}
  {\bibfnamefont {Y.}~\bibnamefont {Lee}}, \bibinfo {author} {\bibfnamefont
  {P.}~\bibnamefont {Rickhaus}}, \bibinfo {author} {\bibfnamefont
  {K.}~\bibnamefont {Watanabe}}, \bibinfo {author} {\bibfnamefont
  {T.}~\bibnamefont {Taniguchi}}, \bibinfo {author} {\bibfnamefont
  {K.}~\bibnamefont {Ensslin}},\ and\ \bibinfo {author} {\bibfnamefont
  {T.}~\bibnamefont {Ihn}},\ }\bibfield  {title} {\bibinfo {title} {{Coupled
  Quantum Dots in Bilayer Graphene}},\ }\href
  {https://doi.org/10.1021/acs.nanolett.8b01859} {\bibfield  {journal}
  {\bibinfo  {journal} {Nano Lett.}\ }\textbf {\bibinfo {volume} {18}},\
  \bibinfo {pages} {5042} (\bibinfo {year} {2018})}\BibitemShut {NoStop}%
\bibitem [{\citenamefont {Ajayi}\ \emph {et~al.}(2017)\citenamefont {Ajayi},
  \citenamefont {Ardelean}, \citenamefont {Shepard}, \citenamefont {Wang},
  \citenamefont {Antony}, \citenamefont {Taniguchi}, \citenamefont {Watanabe},
  \citenamefont {Heinz}, \citenamefont {Strauf}, \citenamefont {Zhu},\ and\
  \citenamefont {Hone}}]{Ajayi2017Jul}%
  \BibitemOpen
  \bibfield  {author} {\bibinfo {author} {\bibfnamefont {O.~A.}\ \bibnamefont
  {Ajayi}}, \bibinfo {author} {\bibfnamefont {J.~V.}\ \bibnamefont {Ardelean}},
  \bibinfo {author} {\bibfnamefont {G.~D.}\ \bibnamefont {Shepard}}, \bibinfo
  {author} {\bibfnamefont {J.}~\bibnamefont {Wang}}, \bibinfo {author}
  {\bibfnamefont {A.}~\bibnamefont {Antony}}, \bibinfo {author} {\bibfnamefont
  {T.}~\bibnamefont {Taniguchi}}, \bibinfo {author} {\bibfnamefont
  {K.}~\bibnamefont {Watanabe}}, \bibinfo {author} {\bibfnamefont {T.~F.}\
  \bibnamefont {Heinz}}, \bibinfo {author} {\bibfnamefont {S.}~\bibnamefont
  {Strauf}}, \bibinfo {author} {\bibfnamefont {X.-Y.}\ \bibnamefont {Zhu}},\
  and\ \bibinfo {author} {\bibfnamefont {J.~C.}\ \bibnamefont {Hone}},\
  }\bibfield  {title} {\bibinfo {title} {{Approaching the intrinsic
  photoluminescence linewidth in transition metal dichalcogenide monolayers}},\
  }\href {https://doi.org/10.1088/2053-1583/aa6aa1} {\bibfield  {journal}
  {\bibinfo  {journal} {2D Mater.}\ }\textbf {\bibinfo {volume} {4}},\ \bibinfo
  {pages} {031011} (\bibinfo {year} {2017})}\BibitemShut {NoStop}%
\bibitem [{\citenamefont {Raja}\ \emph {et~al.}(2019)\citenamefont {Raja},
  \citenamefont {Waldecker}, \citenamefont {Zipfel}, \citenamefont {Cho},
  \citenamefont {Brem}, \citenamefont {Ziegler}, \citenamefont {Kulig},
  \citenamefont {Taniguchi}, \citenamefont {Watanabe}, \citenamefont {Malic},
  \citenamefont {Heinz}, \citenamefont {Berkelbach},\ and\ \citenamefont
  {Chernikov}}]{Raja2019Sep}%
  \BibitemOpen
  \bibfield  {author} {\bibinfo {author} {\bibfnamefont {A.}~\bibnamefont
  {Raja}}, \bibinfo {author} {\bibfnamefont {L.}~\bibnamefont {Waldecker}},
  \bibinfo {author} {\bibfnamefont {J.}~\bibnamefont {Zipfel}}, \bibinfo
  {author} {\bibfnamefont {Y.}~\bibnamefont {Cho}}, \bibinfo {author}
  {\bibfnamefont {S.}~\bibnamefont {Brem}}, \bibinfo {author} {\bibfnamefont
  {J.~D.}\ \bibnamefont {Ziegler}}, \bibinfo {author} {\bibfnamefont
  {M.}~\bibnamefont {Kulig}}, \bibinfo {author} {\bibfnamefont
  {T.}~\bibnamefont {Taniguchi}}, \bibinfo {author} {\bibfnamefont
  {K.}~\bibnamefont {Watanabe}}, \bibinfo {author} {\bibfnamefont
  {E.}~\bibnamefont {Malic}}, \bibinfo {author} {\bibfnamefont {T.~F.}\
  \bibnamefont {Heinz}}, \bibinfo {author} {\bibfnamefont {T.~C.}\ \bibnamefont
  {Berkelbach}},\ and\ \bibinfo {author} {\bibfnamefont {A.}~\bibnamefont
  {Chernikov}},\ }\bibfield  {title} {\bibinfo {title} {{Dielectric disorder in
  two-dimensional materials}},\ }\href
  {https://doi.org/10.1038/s41565-019-0520-0} {\bibfield  {journal} {\bibinfo
  {journal} {Nat. Nanotechnol.}\ }\textbf {\bibinfo {volume} {14}},\ \bibinfo
  {pages} {832} (\bibinfo {year} {2019})}\BibitemShut {NoStop}%
\bibitem [{\citenamefont {Ersfeld}\ \emph {et~al.}(2020)\citenamefont
  {Ersfeld}, \citenamefont {Volmer}, \citenamefont {Rathmann}, \citenamefont
  {Kotewitz}, \citenamefont {Heithoff}, \citenamefont {Lohmann}, \citenamefont
  {Yang}, \citenamefont {Watanabe}, \citenamefont {Taniguchi}, \citenamefont
  {Bartels}, \citenamefont {Shi}, \citenamefont {Stampfer},\ and\ \citenamefont
  {Beschoten}}]{Ersfeld2020May}%
  \BibitemOpen
  \bibfield  {author} {\bibinfo {author} {\bibfnamefont {M.}~\bibnamefont
  {Ersfeld}}, \bibinfo {author} {\bibfnamefont {F.}~\bibnamefont {Volmer}},
  \bibinfo {author} {\bibfnamefont {L.}~\bibnamefont {Rathmann}}, \bibinfo
  {author} {\bibfnamefont {L.}~\bibnamefont {Kotewitz}}, \bibinfo {author}
  {\bibfnamefont {M.}~\bibnamefont {Heithoff}}, \bibinfo {author}
  {\bibfnamefont {M.}~\bibnamefont {Lohmann}}, \bibinfo {author} {\bibfnamefont
  {B.}~\bibnamefont {Yang}}, \bibinfo {author} {\bibfnamefont {K.}~\bibnamefont
  {Watanabe}}, \bibinfo {author} {\bibfnamefont {T.}~\bibnamefont {Taniguchi}},
  \bibinfo {author} {\bibfnamefont {L.}~\bibnamefont {Bartels}}, \bibinfo
  {author} {\bibfnamefont {J.}~\bibnamefont {Shi}}, \bibinfo {author}
  {\bibfnamefont {C.}~\bibnamefont {Stampfer}},\ and\ \bibinfo {author}
  {\bibfnamefont {B.}~\bibnamefont {Beschoten}},\ }\bibfield  {title} {\bibinfo
  {title} {{Unveiling Valley Lifetimes of Free Charge Carriers in Monolayer
  WSe$_2$}},\ }\href {https://doi.org/10.1021/acs.nanolett.9b05138} {\bibfield
  {journal} {\bibinfo  {journal} {Nano Lett.}\ }\textbf {\bibinfo {volume}
  {20}},\ \bibinfo {pages} {3147} (\bibinfo {year} {2020})}\BibitemShut
  {NoStop}%
\bibitem [{\citenamefont {Meng}\ \emph {et~al.}(2019)\citenamefont {Meng},
  \citenamefont {Wang}, \citenamefont {Cheng}, \citenamefont {Gao},\ and\
  \citenamefont {Zhang}}]{Meng2019Feb}%
  \BibitemOpen
  \bibfield  {author} {\bibinfo {author} {\bibfnamefont {J.}~\bibnamefont
  {Meng}}, \bibinfo {author} {\bibfnamefont {D.}~\bibnamefont {Wang}}, \bibinfo
  {author} {\bibfnamefont {L.}~\bibnamefont {Cheng}}, \bibinfo {author}
  {\bibfnamefont {M.}~\bibnamefont {Gao}},\ and\ \bibinfo {author}
  {\bibfnamefont {X.}~\bibnamefont {Zhang}},\ }\bibfield  {title} {\bibinfo
  {title} {{Recent progress in synthesis, properties, and applications of
  hexagonal boron nitride-based heterostructures}},\ }\href
  {https://doi.org/10.1088/1361-6528/aaf301} {\bibfield  {journal} {\bibinfo
  {journal} {Nanotechnology}\ }\textbf {\bibinfo {volume} {30}},\ \bibinfo
  {pages} {074003.} (\bibinfo {year} {2019})}\BibitemShut {NoStop}%
\bibitem [{\citenamefont {Maestre}\ \emph {et~al.}(2021)\citenamefont
  {Maestre}, \citenamefont {Toury}, \citenamefont {Steyer}, \citenamefont
  {Garnier},\ and\ \citenamefont {Journet}}]{Maestre2021Oct}%
  \BibitemOpen
  \bibfield  {author} {\bibinfo {author} {\bibfnamefont {C.}~\bibnamefont
  {Maestre}}, \bibinfo {author} {\bibfnamefont {B.}~\bibnamefont {Toury}},
  \bibinfo {author} {\bibfnamefont {P.}~\bibnamefont {Steyer}}, \bibinfo
  {author} {\bibfnamefont {V.}~\bibnamefont {Garnier}},\ and\ \bibinfo {author}
  {\bibfnamefont {C.}~\bibnamefont {Journet}},\ }\bibfield  {title} {\bibinfo
  {title} {{Hexagonal boron nitride: a review on selfstanding crystals
  synthesis towards 2D nanosheets}},\ }\href
  {https://doi.org/10.1088/2515-7639/ac2b87} {\bibfield  {journal} {\bibinfo
  {journal} {J. Phys.: Mater.}\ }\textbf {\bibinfo {volume} {4}},\ \bibinfo
  {pages} {044018} (\bibinfo {year} {2021})}\BibitemShut {NoStop}%
\bibitem [{\citenamefont {Taniguchi}\ and\ \citenamefont
  {Watanabe}(2007)}]{Taniguchi2007May}%
  \BibitemOpen
  \bibfield  {author} {\bibinfo {author} {\bibfnamefont {T.}~\bibnamefont
  {Taniguchi}}\ and\ \bibinfo {author} {\bibfnamefont {K.}~\bibnamefont
  {Watanabe}},\ }\bibfield  {title} {\bibinfo {title} {{Synthesis of
  high-purity boron nitride single crystals under high pressure by using
  Ba{\textendash}BN solvent}},\ }\href
  {https://doi.org/10.1016/j.jcrysgro.2006.12.061} {\bibfield  {journal}
  {\bibinfo  {journal} {J. Cryst. Growth}\ }\textbf {\bibinfo {volume} {303}},\
  \bibinfo {pages} {525} (\bibinfo {year} {2007})}\BibitemShut {NoStop}%
\bibitem [{\citenamefont {Zastrow}(2019)}]{Zastrow2019Aug}%
  \BibitemOpen
  \bibfield  {author} {\bibinfo {author} {\bibfnamefont {M.}~\bibnamefont
  {Zastrow}},\ }\bibfield  {title} {\bibinfo {title} {{Meet the crystal growers
  who sparked a revolution in graphene electronics}},\ }\href
  {https://doi.org/10.1038/d41586-019-02472-0} {\bibfield  {journal} {\bibinfo
  {journal} {Nature}\ }\textbf {\bibinfo {volume} {572}},\ \bibinfo {pages}
  {429} (\bibinfo {year} {2019})}\BibitemShut {NoStop}%
\bibitem [{\citenamefont {Onodera}\ \emph {et~al.}(2019)\citenamefont
  {Onodera}, \citenamefont {Watanabe}, \citenamefont {Isayama}, \citenamefont
  {Arai}, \citenamefont {Masubuchi}, \citenamefont {Moriya}, \citenamefont
  {Taniguchi},\ and\ \citenamefont {Machida}}]{Onodera2019Oct}%
  \BibitemOpen
  \bibfield  {author} {\bibinfo {author} {\bibfnamefont {M.}~\bibnamefont
  {Onodera}}, \bibinfo {author} {\bibfnamefont {K.}~\bibnamefont {Watanabe}},
  \bibinfo {author} {\bibfnamefont {M.}~\bibnamefont {Isayama}}, \bibinfo
  {author} {\bibfnamefont {M.}~\bibnamefont {Arai}}, \bibinfo {author}
  {\bibfnamefont {S.}~\bibnamefont {Masubuchi}}, \bibinfo {author}
  {\bibfnamefont {R.}~\bibnamefont {Moriya}}, \bibinfo {author} {\bibfnamefont
  {T.}~\bibnamefont {Taniguchi}},\ and\ \bibinfo {author} {\bibfnamefont
  {T.}~\bibnamefont {Machida}},\ }\bibfield  {title} {\bibinfo {title}
  {{Carbon-Rich Domain in Hexagonal Boron Nitride: Carrier Mobility Degradation
  and Anomalous Bending of the Landau Fan Diagram in Adjacent Graphene}},\
  }\href {https://doi.org/10.1021/acs.nanolett.9b02879} {\bibfield  {journal}
  {\bibinfo  {journal} {Nano Lett.}\ }\textbf {\bibinfo {volume} {19}},\
  \bibinfo {pages} {7282} (\bibinfo {year} {2019})}\BibitemShut {NoStop}%
\bibitem [{\citenamefont {Onodera}\ \emph {et~al.}(2020)\citenamefont
  {Onodera}, \citenamefont {Taniguchi}, \citenamefont {Watanabe}, \citenamefont
  {Isayama}, \citenamefont {Masubuchi}, \citenamefont {Moriya},\ and\
  \citenamefont {Machida}}]{Onodera2020Jan}%
  \BibitemOpen
  \bibfield  {author} {\bibinfo {author} {\bibfnamefont {M.}~\bibnamefont
  {Onodera}}, \bibinfo {author} {\bibfnamefont {T.}~\bibnamefont {Taniguchi}},
  \bibinfo {author} {\bibfnamefont {K.}~\bibnamefont {Watanabe}}, \bibinfo
  {author} {\bibfnamefont {M.}~\bibnamefont {Isayama}}, \bibinfo {author}
  {\bibfnamefont {S.}~\bibnamefont {Masubuchi}}, \bibinfo {author}
  {\bibfnamefont {R.}~\bibnamefont {Moriya}},\ and\ \bibinfo {author}
  {\bibfnamefont {T.}~\bibnamefont {Machida}},\ }\bibfield  {title} {\bibinfo
  {title} {{Hexagonal Boron Nitride Synthesized at Atmospheric Pressure Using
  Metal Alloy Solvents: Evaluation as a Substrate for 2D Materials}},\ }\href
  {https://doi.org/10.1021/acs.nanolett.9b04641} {\bibfield  {journal}
  {\bibinfo  {journal} {Nano Lett.}\ }\textbf {\bibinfo {volume} {20}},\
  \bibinfo {pages} {735} (\bibinfo {year} {2020})}\BibitemShut {NoStop}%
\bibitem [{\citenamefont {Kubota}\ \emph
  {et~al.}(2007{\natexlab{a}})\citenamefont {Kubota}, \citenamefont {Watanabe},
  \citenamefont {Tsuda},\ and\ \citenamefont {Taniguchi}}]{Kubota2007Aug}%
  \BibitemOpen
  \bibfield  {author} {\bibinfo {author} {\bibfnamefont {Y.}~\bibnamefont
  {Kubota}}, \bibinfo {author} {\bibfnamefont {K.}~\bibnamefont {Watanabe}},
  \bibinfo {author} {\bibfnamefont {O.}~\bibnamefont {Tsuda}},\ and\ \bibinfo
  {author} {\bibfnamefont {T.}~\bibnamefont {Taniguchi}},\ }\bibfield  {title}
  {\bibinfo {title} {{Deep Ultraviolet Light-Emitting Hexagonal Boron Nitride
  Synthesized at Atmospheric Pressure}},\ }\href
  {https://doi.org/10.1126/science.1144216} {\bibfield  {journal} {\bibinfo
  {journal} {Science}\ }\textbf {\bibinfo {volume} {317}},\ \bibinfo {pages}
  {932} (\bibinfo {year} {2007}{\natexlab{a}})}\BibitemShut {NoStop}%
\bibitem [{\citenamefont {Kubota}\ \emph {et~al.}(2008)\citenamefont {Kubota},
  \citenamefont {Watanabe}, \citenamefont {Tsuda},\ and\ \citenamefont
  {Taniguchi}}]{Kubota2008Mar}%
  \BibitemOpen
  \bibfield  {author} {\bibinfo {author} {\bibfnamefont {Y.}~\bibnamefont
  {Kubota}}, \bibinfo {author} {\bibfnamefont {K.}~\bibnamefont {Watanabe}},
  \bibinfo {author} {\bibfnamefont {O.}~\bibnamefont {Tsuda}},\ and\ \bibinfo
  {author} {\bibfnamefont {T.}~\bibnamefont {Taniguchi}},\ }\bibfield  {title}
  {\bibinfo {title} {{Hexagonal Boron Nitride Single Crystal Growth at
  Atmospheric Pressure Using Ni{-}Cr Solvent}},\ }\href
  {https://doi.org/10.1021/cm7028382} {\bibfield  {journal} {\bibinfo
  {journal} {Chem. Mater.}\ }\textbf {\bibinfo {volume} {20}},\ \bibinfo
  {pages} {1661} (\bibinfo {year} {2008})}\BibitemShut {NoStop}%
\bibitem [{\citenamefont {Kubota}\ \emph
  {et~al.}(2007{\natexlab{b}})\citenamefont {Kubota}, \citenamefont
  {Watanabe},\ and\ \citenamefont {Taniguchi}}]{Kubota2007Jan}%
  \BibitemOpen
  \bibfield  {author} {\bibinfo {author} {\bibfnamefont {Y.}~\bibnamefont
  {Kubota}}, \bibinfo {author} {\bibfnamefont {K.}~\bibnamefont {Watanabe}},\
  and\ \bibinfo {author} {\bibfnamefont {T.}~\bibnamefont {Taniguchi}},\
  }\bibfield  {title} {\bibinfo {title} {{Synthesis of Cubic and Hexagonal
  Boron Nitrides by Using Ni Solvent under High Pressure}},\ }\href
  {https://doi.org/10.1143/JJAP.46.311} {\bibfield  {journal} {\bibinfo
  {journal} {Jpn. J. Appl. Phys.}\ }\textbf {\bibinfo {volume} {46}},\ \bibinfo
  {pages} {311} (\bibinfo {year} {2007}{\natexlab{b}})}\BibitemShut {NoStop}%
\bibitem [{\citenamefont {Hoffman}\ \emph {et~al.}(2014)\citenamefont
  {Hoffman}, \citenamefont {Clubine}, \citenamefont {Zhang}, \citenamefont
  {Snow},\ and\ \citenamefont {Edgar}}]{Hoffman2014May}%
  \BibitemOpen
  \bibfield  {author} {\bibinfo {author} {\bibfnamefont {T.~B.}\ \bibnamefont
  {Hoffman}}, \bibinfo {author} {\bibfnamefont {B.}~\bibnamefont {Clubine}},
  \bibinfo {author} {\bibfnamefont {Y.}~\bibnamefont {Zhang}}, \bibinfo
  {author} {\bibfnamefont {K.}~\bibnamefont {Snow}},\ and\ \bibinfo {author}
  {\bibfnamefont {J.~H.}\ \bibnamefont {Edgar}},\ }\bibfield  {title} {\bibinfo
  {title} {{Optimization of Ni{\textendash}Cr flux growth for hexagonal boron
  nitride single crystals}},\ }\href
  {https://doi.org/10.1016/j.jcrysgro.2013.09.030} {\bibfield  {journal}
  {\bibinfo  {journal} {J. Cryst. Growth}\ }\textbf {\bibinfo {volume} {393}},\
  \bibinfo {pages} {114} (\bibinfo {year} {2014})}\BibitemShut {NoStop}%
\bibitem [{\citenamefont {Edgar}\ \emph {et~al.}(2014)\citenamefont {Edgar},
  \citenamefont {Hoffman}, \citenamefont {Clubine}, \citenamefont {Currie},
  \citenamefont {Du}, \citenamefont {Lin},\ and\ \citenamefont
  {Jiang}}]{Edgar2014Oct}%
  \BibitemOpen
  \bibfield  {author} {\bibinfo {author} {\bibfnamefont {J.~H.}\ \bibnamefont
  {Edgar}}, \bibinfo {author} {\bibfnamefont {T.~B.}\ \bibnamefont {Hoffman}},
  \bibinfo {author} {\bibfnamefont {B.}~\bibnamefont {Clubine}}, \bibinfo
  {author} {\bibfnamefont {M.}~\bibnamefont {Currie}}, \bibinfo {author}
  {\bibfnamefont {X.~Z.}\ \bibnamefont {Du}}, \bibinfo {author} {\bibfnamefont
  {J.~Y.}\ \bibnamefont {Lin}},\ and\ \bibinfo {author} {\bibfnamefont {H.~X.}\
  \bibnamefont {Jiang}},\ }\bibfield  {title} {\bibinfo {title}
  {{Characterization of bulk hexagonal boron nitride single crystals grown by
  the metal flux technique}},\ }\href
  {https://doi.org/10.1016/j.jcrysgro.2014.06.006} {\bibfield  {journal}
  {\bibinfo  {journal} {J. Cryst. Growth}\ }\textbf {\bibinfo {volume} {403}},\
  \bibinfo {pages} {110} (\bibinfo {year} {2014})}\BibitemShut {NoStop}%
\bibitem [{\citenamefont {Liu}\ \emph {et~al.}(2018)\citenamefont {Liu},
  \citenamefont {He}, \citenamefont {Xue}, \citenamefont {Li}, \citenamefont
  {Liu},\ and\ \citenamefont {Edgar}}]{Liu2018Sep}%
  \BibitemOpen
  \bibfield  {author} {\bibinfo {author} {\bibfnamefont {S.}~\bibnamefont
  {Liu}}, \bibinfo {author} {\bibfnamefont {R.}~\bibnamefont {He}}, \bibinfo
  {author} {\bibfnamefont {L.}~\bibnamefont {Xue}}, \bibinfo {author}
  {\bibfnamefont {J.}~\bibnamefont {Li}}, \bibinfo {author} {\bibfnamefont
  {B.}~\bibnamefont {Liu}},\ and\ \bibinfo {author} {\bibfnamefont {J.~H.}\
  \bibnamefont {Edgar}},\ }\bibfield  {title} {\bibinfo {title} {{Single
  Crystal Growth of Millimeter-Sized Monoisotopic Hexagonal Boron Nitride}},\
  }\href {https://doi.org/10.1021/acs.chemmater.8b02589} {\bibfield  {journal}
  {\bibinfo  {journal} {Chem. Mater.}\ }\textbf {\bibinfo {volume} {30}},\
  \bibinfo {pages} {6222} (\bibinfo {year} {2018})}\BibitemShut {NoStop}%
\bibitem [{\citenamefont {Li}\ \emph {et~al.}(2021{\natexlab{a}})\citenamefont
  {Li}, \citenamefont {Glaser}, \citenamefont {Elias}, \citenamefont {Ye},
  \citenamefont {Evans}, \citenamefont {Xue}, \citenamefont {Liu},
  \citenamefont {Cassabois}, \citenamefont {Gil}, \citenamefont {Valvin},
  \citenamefont {Pelini}, \citenamefont {Yeats}, \citenamefont {He},
  \citenamefont {Liu},\ and\ \citenamefont {Edgar}}]{li2021Dec}%
  \BibitemOpen
  \bibfield  {author} {\bibinfo {author} {\bibfnamefont {J.}~\bibnamefont
  {Li}}, \bibinfo {author} {\bibfnamefont {E.~R.}\ \bibnamefont {Glaser}},
  \bibinfo {author} {\bibfnamefont {C.}~\bibnamefont {Elias}}, \bibinfo
  {author} {\bibfnamefont {G.}~\bibnamefont {Ye}}, \bibinfo {author}
  {\bibfnamefont {D.}~\bibnamefont {Evans}}, \bibinfo {author} {\bibfnamefont
  {L.}~\bibnamefont {Xue}}, \bibinfo {author} {\bibfnamefont {S.}~\bibnamefont
  {Liu}}, \bibinfo {author} {\bibfnamefont {G.}~\bibnamefont {Cassabois}},
  \bibinfo {author} {\bibfnamefont {B.}~\bibnamefont {Gil}}, \bibinfo {author}
  {\bibfnamefont {P.}~\bibnamefont {Valvin}}, \bibinfo {author} {\bibfnamefont
  {T.}~\bibnamefont {Pelini}}, \bibinfo {author} {\bibfnamefont {A.~L.}\
  \bibnamefont {Yeats}}, \bibinfo {author} {\bibfnamefont {R.}~\bibnamefont
  {He}}, \bibinfo {author} {\bibfnamefont {B.}~\bibnamefont {Liu}},\ and\
  \bibinfo {author} {\bibfnamefont {J.~H.}\ \bibnamefont {Edgar}},\ }\bibfield
  {title} {\bibinfo {title} {{Defect Engineering of Monoisotopic Hexagonal
  Boron Nitride Crystals via Neutron Transmutation Doping}},\ }\href
  {https://doi.org/10.1021/acs.chemmater.1c02849} {\bibfield  {journal}
  {\bibinfo  {journal} {Chem. Mater.}\ }\textbf {\bibinfo {volume} {33}},\
  \bibinfo {pages} {9231} (\bibinfo {year} {2021}{\natexlab{a}})}\BibitemShut
  {NoStop}%
\bibitem [{\citenamefont {Li}\ \emph {et~al.}(2020{\natexlab{a}})\citenamefont
  {Li}, \citenamefont {Elias}, \citenamefont {Ye}, \citenamefont {Evans},
  \citenamefont {Liu}, \citenamefont {He}, \citenamefont {Cassabois},
  \citenamefont {Gil}, \citenamefont {Valvin}, \citenamefont {Liu},\ and\
  \citenamefont {Edgar}}]{Li2020}%
  \BibitemOpen
  \bibfield  {author} {\bibinfo {author} {\bibfnamefont {J.}~\bibnamefont
  {Li}}, \bibinfo {author} {\bibfnamefont {C.}~\bibnamefont {Elias}}, \bibinfo
  {author} {\bibfnamefont {G.}~\bibnamefont {Ye}}, \bibinfo {author}
  {\bibfnamefont {D.}~\bibnamefont {Evans}}, \bibinfo {author} {\bibfnamefont
  {S.}~\bibnamefont {Liu}}, \bibinfo {author} {\bibfnamefont {R.}~\bibnamefont
  {He}}, \bibinfo {author} {\bibfnamefont {G.}~\bibnamefont {Cassabois}},
  \bibinfo {author} {\bibfnamefont {B.}~\bibnamefont {Gil}}, \bibinfo {author}
  {\bibfnamefont {P.}~\bibnamefont {Valvin}}, \bibinfo {author} {\bibfnamefont
  {B.}~\bibnamefont {Liu}},\ and\ \bibinfo {author} {\bibfnamefont {J.~H.}\
  \bibnamefont {Edgar}},\ }\bibfield  {title} {\bibinfo {title} {{Single
  crystal growth of monoisotopic hexagonal boron nitride from a
  Fe{\textendash}Cr flux}},\ }\href {https://doi.org/10.1039/D0TC02143A}
  {\bibfield  {journal} {\bibinfo  {journal} {J. Mater. Chem. C}\ }\textbf
  {\bibinfo {volume} {8}},\ \bibinfo {pages} {9931} (\bibinfo {year}
  {2020}{\natexlab{a}})}\BibitemShut {NoStop}%
\bibitem [{\citenamefont {Zhang}\ \emph {et~al.}(2019)\citenamefont {Zhang},
  \citenamefont {Xu}, \citenamefont {Zhao}, \citenamefont {Shao},\ and\
  \citenamefont {Wan}}]{Zhang2019Nov}%
  \BibitemOpen
  \bibfield  {author} {\bibinfo {author} {\bibfnamefont {S.-Y.}\ \bibnamefont
  {Zhang}}, \bibinfo {author} {\bibfnamefont {K.}~\bibnamefont {Xu}}, \bibinfo
  {author} {\bibfnamefont {X.-K.}\ \bibnamefont {Zhao}}, \bibinfo {author}
  {\bibfnamefont {Z.-Y.}\ \bibnamefont {Shao}},\ and\ \bibinfo {author}
  {\bibfnamefont {N.}~\bibnamefont {Wan}},\ }\bibfield  {title} {\bibinfo
  {title} {{Improved hBN Single-Crystal Growth by Adding Carbon in the Metal
  Flux}},\ }\href {https://doi.org/10.1021/acs.cgd.9b00712} {\bibfield
  {journal} {\bibinfo  {journal} {Cryst. Growth Des.}\ }\textbf {\bibinfo
  {volume} {19}},\ \bibinfo {pages} {6252} (\bibinfo {year}
  {2019})}\BibitemShut {NoStop}%
\bibitem [{\citenamefont {Cao}\ \emph {et~al.}(2022)\citenamefont {Cao},
  \citenamefont {Tian}, \citenamefont {Zhang}, \citenamefont {Hu},
  \citenamefont {Wan},\ and\ \citenamefont {Lin}}]{Cao2022Aug}%
  \BibitemOpen
  \bibfield  {author} {\bibinfo {author} {\bibfnamefont {J.}~\bibnamefont
  {Cao}}, \bibinfo {author} {\bibfnamefont {M.}~\bibnamefont {Tian}}, \bibinfo
  {author} {\bibfnamefont {S.}~\bibnamefont {Zhang}}, \bibinfo {author}
  {\bibfnamefont {W.}~\bibnamefont {Hu}}, \bibinfo {author} {\bibfnamefont
  {N.}~\bibnamefont {Wan}},\ and\ \bibinfo {author} {\bibfnamefont
  {T.}~\bibnamefont {Lin}},\ }\bibfield  {title} {\bibinfo {title}
  {{Carbon-related defect control of bulk hBN single crystals growth by
  atmospheric-pressure metal-flux-based fusion synthesis}},\ }\href
  {https://doi.org/10.1007/s10853-022-07548-3} {\bibfield  {journal} {\bibinfo
  {journal} {J. Mater. Sci.}\ }\textbf {\bibinfo {volume} {57}},\ \bibinfo
  {pages} {14668} (\bibinfo {year} {2022})}\BibitemShut {NoStop}%
\bibitem [{\citenamefont {Li}\ \emph {et~al.}(2020{\natexlab{b}})\citenamefont
  {Li}, \citenamefont {Yuan}, \citenamefont {Elias}, \citenamefont {Wang},
  \citenamefont {Zhang}, \citenamefont {Ye}, \citenamefont {Huang},
  \citenamefont {Kuball}, \citenamefont {Eda}, \citenamefont {Redwing},
  \citenamefont {He}, \citenamefont {Cassabois}, \citenamefont {Gil},
  \citenamefont {Valvin}, \citenamefont {Pelini}, \citenamefont {Liu},\ and\
  \citenamefont {Edgar}}]{Li2020Jun}%
  \BibitemOpen
  \bibfield  {author} {\bibinfo {author} {\bibfnamefont {J.}~\bibnamefont
  {Li}}, \bibinfo {author} {\bibfnamefont {C.}~\bibnamefont {Yuan}}, \bibinfo
  {author} {\bibfnamefont {C.}~\bibnamefont {Elias}}, \bibinfo {author}
  {\bibfnamefont {J.}~\bibnamefont {Wang}}, \bibinfo {author} {\bibfnamefont
  {X.}~\bibnamefont {Zhang}}, \bibinfo {author} {\bibfnamefont
  {G.}~\bibnamefont {Ye}}, \bibinfo {author} {\bibfnamefont {C.}~\bibnamefont
  {Huang}}, \bibinfo {author} {\bibfnamefont {M.}~\bibnamefont {Kuball}},
  \bibinfo {author} {\bibfnamefont {G.}~\bibnamefont {Eda}}, \bibinfo {author}
  {\bibfnamefont {J.~M.}\ \bibnamefont {Redwing}}, \bibinfo {author}
  {\bibfnamefont {R.}~\bibnamefont {He}}, \bibinfo {author} {\bibfnamefont
  {G.}~\bibnamefont {Cassabois}}, \bibinfo {author} {\bibfnamefont
  {B.}~\bibnamefont {Gil}}, \bibinfo {author} {\bibfnamefont {P.}~\bibnamefont
  {Valvin}}, \bibinfo {author} {\bibfnamefont {T.}~\bibnamefont {Pelini}},
  \bibinfo {author} {\bibfnamefont {B.}~\bibnamefont {Liu}},\ and\ \bibinfo
  {author} {\bibfnamefont {J.~H.}\ \bibnamefont {Edgar}},\ }\bibfield  {title}
  {\bibinfo {title} {{Hexagonal Boron Nitride Single Crystal Growth from
  Solution with a Temperature Gradient}},\ }\href
  {https://doi.org/10.1021/acs.chemmater.0c00830} {\bibfield  {journal}
  {\bibinfo  {journal} {Chem. Mater.}\ }\textbf {\bibinfo {volume} {32}},\
  \bibinfo {pages} {5066} (\bibinfo {year} {2020}{\natexlab{b}})}\BibitemShut
  {NoStop}%
\bibitem [{\citenamefont {Li}\ \emph {et~al.}(2021{\natexlab{b}})\citenamefont
  {Li}, \citenamefont {Wang}, \citenamefont {Zhang}, \citenamefont {Elias},
  \citenamefont {Ye}, \citenamefont {Evans}, \citenamefont {Eda}, \citenamefont
  {Redwing}, \citenamefont {Cassabois}, \citenamefont {Gil}, \citenamefont
  {Valvin}, \citenamefont {He}, \citenamefont {Liu},\ and\ \citenamefont
  {Edgar}}]{Li2021Apr}%
  \BibitemOpen
  \bibfield  {author} {\bibinfo {author} {\bibfnamefont {J.}~\bibnamefont
  {Li}}, \bibinfo {author} {\bibfnamefont {J.}~\bibnamefont {Wang}}, \bibinfo
  {author} {\bibfnamefont {X.}~\bibnamefont {Zhang}}, \bibinfo {author}
  {\bibfnamefont {C.}~\bibnamefont {Elias}}, \bibinfo {author} {\bibfnamefont
  {G.}~\bibnamefont {Ye}}, \bibinfo {author} {\bibfnamefont {D.}~\bibnamefont
  {Evans}}, \bibinfo {author} {\bibfnamefont {G.}~\bibnamefont {Eda}}, \bibinfo
  {author} {\bibfnamefont {J.~M.}\ \bibnamefont {Redwing}}, \bibinfo {author}
  {\bibfnamefont {G.}~\bibnamefont {Cassabois}}, \bibinfo {author}
  {\bibfnamefont {B.}~\bibnamefont {Gil}}, \bibinfo {author} {\bibfnamefont
  {P.}~\bibnamefont {Valvin}}, \bibinfo {author} {\bibfnamefont
  {R.}~\bibnamefont {He}}, \bibinfo {author} {\bibfnamefont {B.}~\bibnamefont
  {Liu}},\ and\ \bibinfo {author} {\bibfnamefont {J.~H.}\ \bibnamefont
  {Edgar}},\ }\bibfield  {title} {\bibinfo {title} {{Hexagonal Boron Nitride
  Crystal Growth from Iron, a Single Component Flux}},\ }\href
  {https://doi.org/10.1021/acsnano.1c00115} {\bibfield  {journal} {\bibinfo
  {journal} {ACS Nano}\ }\textbf {\bibinfo {volume} {15}},\ \bibinfo {pages}
  {7032} (\bibinfo {year} {2021}{\natexlab{b}})}\BibitemShut {NoStop}%
\bibitem [{\citenamefont {Naclerio}\ and\ \citenamefont
  {Kidambi}(2023)}]{Naclerio2023Feb}%
  \BibitemOpen
  \bibfield  {author} {\bibinfo {author} {\bibfnamefont {A.~E.}\ \bibnamefont
  {Naclerio}}\ and\ \bibinfo {author} {\bibfnamefont {P.~R.}\ \bibnamefont
  {Kidambi}},\ }\bibfield  {title} {\bibinfo {title} {{A Review of Scalable
  Hexagonal Boron Nitride (h-BN) Synthesis for Present and Future
  Applications}},\ }\href {https://doi.org/10.1002/adma.202207374} {\bibfield
  {journal} {\bibinfo  {journal} {Adv. Mater.}\ }\textbf {\bibinfo {volume}
  {35}},\ \bibinfo {pages} {2207374} (\bibinfo {year} {2023})}\BibitemShut
  {NoStop}%
\bibitem [{\citenamefont {Liu}\ \emph {et~al.}(2017)\citenamefont {Liu},
  \citenamefont {He}, \citenamefont {Ye}, \citenamefont {Du}, \citenamefont
  {Lin}, \citenamefont {Jiang}, \citenamefont {Liu},\ and\ \citenamefont
  {Edgar}}]{Liu2017Sep}%
  \BibitemOpen
  \bibfield  {author} {\bibinfo {author} {\bibfnamefont {S.}~\bibnamefont
  {Liu}}, \bibinfo {author} {\bibfnamefont {R.}~\bibnamefont {He}}, \bibinfo
  {author} {\bibfnamefont {Z.}~\bibnamefont {Ye}}, \bibinfo {author}
  {\bibfnamefont {X.}~\bibnamefont {Du}}, \bibinfo {author} {\bibfnamefont
  {J.}~\bibnamefont {Lin}}, \bibinfo {author} {\bibfnamefont {H.}~\bibnamefont
  {Jiang}}, \bibinfo {author} {\bibfnamefont {B.}~\bibnamefont {Liu}},\ and\
  \bibinfo {author} {\bibfnamefont {J.~H.}\ \bibnamefont {Edgar}},\ }\bibfield
  {title} {\bibinfo {title} {{Large-Scale Growth of High-Quality Hexagonal
  Boron Nitride Crystals at Atmospheric Pressure from an Fe{\textendash}Cr
  Flux}},\ }\href {https://doi.org/10.1021/acs.cgd.7b00871} {\bibfield
  {journal} {\bibinfo  {journal} {Cryst. Growth Des.}\ }\textbf {\bibinfo
  {volume} {17}},\ \bibinfo {pages} {4932} (\bibinfo {year}
  {2017})}\BibitemShut {NoStop}%
\bibitem [{\citenamefont {Li}\ \emph {et~al.}(2021{\natexlab{c}})\citenamefont
  {Li}, \citenamefont {Wen}, \citenamefont {Tan}, \citenamefont {Li},
  \citenamefont {Li}, \citenamefont {Huang}, \citenamefont {Tian},
  \citenamefont {Yao}, \citenamefont {Liao}, \citenamefont {Yu}, \citenamefont
  {Liu}, \citenamefont {Li}, \citenamefont {Guo}, \citenamefont {Huang},
  \citenamefont {Gao}, \citenamefont {Wang}, \citenamefont {Bai},\ and\
  \citenamefont {Liu}}]{Li2021Jul}%
  \BibitemOpen
  \bibfield  {author} {\bibinfo {author} {\bibfnamefont {Y.}~\bibnamefont
  {Li}}, \bibinfo {author} {\bibfnamefont {X.}~\bibnamefont {Wen}}, \bibinfo
  {author} {\bibfnamefont {C.}~\bibnamefont {Tan}}, \bibinfo {author}
  {\bibfnamefont {N.}~\bibnamefont {Li}}, \bibinfo {author} {\bibfnamefont
  {R.}~\bibnamefont {Li}}, \bibinfo {author} {\bibfnamefont {X.}~\bibnamefont
  {Huang}}, \bibinfo {author} {\bibfnamefont {H.}~\bibnamefont {Tian}},
  \bibinfo {author} {\bibfnamefont {Z.}~\bibnamefont {Yao}}, \bibinfo {author}
  {\bibfnamefont {P.}~\bibnamefont {Liao}}, \bibinfo {author} {\bibfnamefont
  {S.}~\bibnamefont {Yu}}, \bibinfo {author} {\bibfnamefont {S.}~\bibnamefont
  {Liu}}, \bibinfo {author} {\bibfnamefont {Z.}~\bibnamefont {Li}}, \bibinfo
  {author} {\bibfnamefont {J.}~\bibnamefont {Guo}}, \bibinfo {author}
  {\bibfnamefont {Y.}~\bibnamefont {Huang}}, \bibinfo {author} {\bibfnamefont
  {P.}~\bibnamefont {Gao}}, \bibinfo {author} {\bibfnamefont {L.}~\bibnamefont
  {Wang}}, \bibinfo {author} {\bibfnamefont {S.}~\bibnamefont {Bai}},\ and\
  \bibinfo {author} {\bibfnamefont {L.}~\bibnamefont {Liu}},\ }\bibfield
  {title} {\bibinfo {title} {{Synthesis of centimeter-scale high-quality
  polycrystalline hexagonal boron nitride films from Fe fluxes}},\ }\href
  {https://doi.org/10.1039/D1NR02408F} {\bibfield  {journal} {\bibinfo
  {journal} {Nanoscale}\ }\textbf {\bibinfo {volume} {13}},\ \bibinfo {pages}
  {11223} (\bibinfo {year} {2021}{\natexlab{c}})}\BibitemShut {NoStop}%
\bibitem [{\citenamefont {Suk}\ \emph {et~al.}(2011)\citenamefont {Suk},
  \citenamefont {Kitt}, \citenamefont {Magnuson}, \citenamefont {Hao},
  \citenamefont {Ahmed}, \citenamefont {An}, \citenamefont {Swan},
  \citenamefont {Goldberg},\ and\ \citenamefont {Ruoff}}]{Suk2011Sep}%
  \BibitemOpen
  \bibfield  {author} {\bibinfo {author} {\bibfnamefont {J.~W.}\ \bibnamefont
  {Suk}}, \bibinfo {author} {\bibfnamefont {A.}~\bibnamefont {Kitt}}, \bibinfo
  {author} {\bibfnamefont {C.~W.}\ \bibnamefont {Magnuson}}, \bibinfo {author}
  {\bibfnamefont {Y.}~\bibnamefont {Hao}}, \bibinfo {author} {\bibfnamefont
  {S.}~\bibnamefont {Ahmed}}, \bibinfo {author} {\bibfnamefont
  {J.}~\bibnamefont {An}}, \bibinfo {author} {\bibfnamefont {A.~K.}\
  \bibnamefont {Swan}}, \bibinfo {author} {\bibfnamefont {B.~B.}\ \bibnamefont
  {Goldberg}},\ and\ \bibinfo {author} {\bibfnamefont {R.~S.}\ \bibnamefont
  {Ruoff}},\ }\bibfield  {title} {\bibinfo {title} {{Transfer of CVD-Grown
  Monolayer Graphene onto Arbitrary Substrates}},\ }\href
  {https://doi.org/10.1021/nn201207c} {\bibfield  {journal} {\bibinfo
  {journal} {ACS Nano}\ }\textbf {\bibinfo {volume} {5}},\ \bibinfo {pages}
  {6916} (\bibinfo {year} {2011})}\BibitemShut {NoStop}%
\bibitem [{\citenamefont {Schu{\ifmmode\acute{e}\else\'{e}\fi}}\ \emph
  {et~al.}(2016)\citenamefont {Schu{\ifmmode\acute{e}\else\'{e}\fi}},
  \citenamefont {Stenger}, \citenamefont {Fossard}, \citenamefont {Loiseau},\
  and\ \citenamefont {Barjon}}]{Schue2016Dec}%
  \BibitemOpen
  \bibfield  {author} {\bibinfo {author} {\bibfnamefont {L.}~\bibnamefont
  {Schu{\ifmmode\acute{e}\else\'{e}\fi}}}, \bibinfo {author} {\bibfnamefont
  {I.}~\bibnamefont {Stenger}}, \bibinfo {author} {\bibfnamefont
  {F.}~\bibnamefont {Fossard}}, \bibinfo {author} {\bibfnamefont
  {A.}~\bibnamefont {Loiseau}},\ and\ \bibinfo {author} {\bibfnamefont
  {J.}~\bibnamefont {Barjon}},\ }\bibfield  {title} {\bibinfo {title}
  {{Characterization methods dedicated to nanometer-thick hBN layers}},\ }\href
  {https://doi.org/10.1088/2053-1583/4/1/015028} {\bibfield  {journal}
  {\bibinfo  {journal} {2D Mater.}\ }\textbf {\bibinfo {volume} {4}},\ \bibinfo
  {pages} {015028} (\bibinfo {year} {2016})}\BibitemShut {NoStop}%
\bibitem [{\citenamefont {Sonntag}\ \emph {et~al.}(2020)\citenamefont
  {Sonntag}, \citenamefont {Li}, \citenamefont {Plaud}, \citenamefont
  {Loiseau}, \citenamefont {Barjon}, \citenamefont {Edgar},\ and\ \citenamefont
  {Stampfer}}]{Sonntag2020Jun}%
  \BibitemOpen
  \bibfield  {author} {\bibinfo {author} {\bibfnamefont {J.}~\bibnamefont
  {Sonntag}}, \bibinfo {author} {\bibfnamefont {J.}~\bibnamefont {Li}},
  \bibinfo {author} {\bibfnamefont {A.}~\bibnamefont {Plaud}}, \bibinfo
  {author} {\bibfnamefont {A.}~\bibnamefont {Loiseau}}, \bibinfo {author}
  {\bibfnamefont {J.}~\bibnamefont {Barjon}}, \bibinfo {author} {\bibfnamefont
  {J.~H.}\ \bibnamefont {Edgar}},\ and\ \bibinfo {author} {\bibfnamefont
  {C.}~\bibnamefont {Stampfer}},\ }\bibfield  {title} {\bibinfo {title}
  {{Excellent electronic transport in heterostructures of graphene and
  monoisotopic boron-nitride grown at atmospheric pressure}},\ }\href
  {https://doi.org/10.1088/2053-1583/ab89e5} {\bibfield  {journal} {\bibinfo
  {journal} {2D Mater.}\ }\textbf {\bibinfo {volume} {7}},\ \bibinfo {pages}
  {031009} (\bibinfo {year} {2020})}\BibitemShut {NoStop}%
\bibitem [{\citenamefont {Neumann}\ \emph {et~al.}(2015)\citenamefont
  {Neumann}, \citenamefont {Reichardt}, \citenamefont {Venezuela},
  \citenamefont {Dr{\ifmmode\ddot{o}\else\"{o}\fi}geler}, \citenamefont
  {Banszerus}, \citenamefont {Schmitz}, \citenamefont {Watanabe}, \citenamefont
  {Taniguchi}, \citenamefont {Mauri}, \citenamefont {Beschoten}, \citenamefont
  {Rotkin},\ and\ \citenamefont {Stampfer}}]{Neumann2015Sep}%
  \BibitemOpen
  \bibfield  {author} {\bibinfo {author} {\bibfnamefont {C.}~\bibnamefont
  {Neumann}}, \bibinfo {author} {\bibfnamefont {S.}~\bibnamefont {Reichardt}},
  \bibinfo {author} {\bibfnamefont {P.}~\bibnamefont {Venezuela}}, \bibinfo
  {author} {\bibfnamefont {M.}~\bibnamefont
  {Dr{\ifmmode\ddot{o}\else\"{o}\fi}geler}}, \bibinfo {author} {\bibfnamefont
  {L.}~\bibnamefont {Banszerus}}, \bibinfo {author} {\bibfnamefont
  {M.}~\bibnamefont {Schmitz}}, \bibinfo {author} {\bibfnamefont
  {K.}~\bibnamefont {Watanabe}}, \bibinfo {author} {\bibfnamefont
  {T.}~\bibnamefont {Taniguchi}}, \bibinfo {author} {\bibfnamefont
  {F.}~\bibnamefont {Mauri}}, \bibinfo {author} {\bibfnamefont
  {B.}~\bibnamefont {Beschoten}}, \bibinfo {author} {\bibfnamefont {S.~V.}\
  \bibnamefont {Rotkin}},\ and\ \bibinfo {author} {\bibfnamefont
  {C.}~\bibnamefont {Stampfer}},\ }\bibfield  {title} {\bibinfo {title} {{Raman
  spectroscopy as probe of nanometre-scale strain variations in graphene}},\
  }\href {https://doi.org/10.1038/ncomms9429} {\bibfield  {journal} {\bibinfo
  {journal} {Nat. Commun.}\ }\textbf {\bibinfo {volume} {6}},\ \bibinfo {pages}
  {8429} (\bibinfo {year} {2015})}\BibitemShut {NoStop}%
\bibitem [{\citenamefont {Cusc{\ifmmode\acute{o}\else\'{o}\fi}}\ \emph
  {et~al.}(2020)\citenamefont {Cusc{\ifmmode\acute{o}\else\'{o}\fi}},
  \citenamefont {Edgar}, \citenamefont {Liu}, \citenamefont {Li},\ and\
  \citenamefont {Art{\ifmmode\acute{u}\else\'{u}\fi}s}}]{Cusco2020Apr}%
  \BibitemOpen
  \bibfield  {author} {\bibinfo {author} {\bibfnamefont {R.}~\bibnamefont
  {Cusc{\ifmmode\acute{o}\else\'{o}\fi}}}, \bibinfo {author} {\bibfnamefont
  {J.~H.}\ \bibnamefont {Edgar}}, \bibinfo {author} {\bibfnamefont
  {S.}~\bibnamefont {Liu}}, \bibinfo {author} {\bibfnamefont {J.}~\bibnamefont
  {Li}},\ and\ \bibinfo {author} {\bibfnamefont {L.}~\bibnamefont
  {Art{\ifmmode\acute{u}\else\'{u}\fi}s}},\ }\bibfield  {title} {\bibinfo
  {title} {{Isotopic Disorder: The Prevailing Mechanism in Limiting the Phonon
  Lifetime in Hexagonal BN}},\ }\href
  {https://doi.org/10.1103/PhysRevLett.124.167402} {\bibfield  {journal}
  {\bibinfo  {journal} {Phys. Rev. Lett.}\ }\textbf {\bibinfo {volume} {124}},\
  \bibinfo {pages} {167402} (\bibinfo {year} {2020})}\BibitemShut {NoStop}%
\bibitem [{\citenamefont {Bisswanger}\ \emph {et~al.}(2022)\citenamefont
  {Bisswanger}, \citenamefont {Winter}, \citenamefont {Schmidt}, \citenamefont
  {Volmer}, \citenamefont {Watanabe}, \citenamefont {Taniguchi}, \citenamefont
  {Stampfer},\ and\ \citenamefont {Beschoten}}]{Bisswanger2022Jun}%
  \BibitemOpen
  \bibfield  {author} {\bibinfo {author} {\bibfnamefont {T.}~\bibnamefont
  {Bisswanger}}, \bibinfo {author} {\bibfnamefont {Z.}~\bibnamefont {Winter}},
  \bibinfo {author} {\bibfnamefont {A.}~\bibnamefont {Schmidt}}, \bibinfo
  {author} {\bibfnamefont {F.}~\bibnamefont {Volmer}}, \bibinfo {author}
  {\bibfnamefont {K.}~\bibnamefont {Watanabe}}, \bibinfo {author}
  {\bibfnamefont {T.}~\bibnamefont {Taniguchi}}, \bibinfo {author}
  {\bibfnamefont {C.}~\bibnamefont {Stampfer}},\ and\ \bibinfo {author}
  {\bibfnamefont {B.}~\bibnamefont {Beschoten}},\ }\bibfield  {title} {\bibinfo
  {title} {{CVD Bilayer Graphene Spin Valves with 26 {$\mu$}m Spin Diffusion
  Length at Room Temperature}},\ }\href
  {https://doi.org/10.1021/acs.nanolett.2c01119} {\bibfield  {journal}
  {\bibinfo  {journal} {Nano Lett.}\ }\textbf {\bibinfo {volume} {22}},\
  \bibinfo {pages} {4949} (\bibinfo {year} {2022})}\BibitemShut {NoStop}%
\bibitem [{\citenamefont {Couto}\ \emph {et~al.}(2014)\citenamefont {Couto},
  \citenamefont {Costanzo}, \citenamefont {Engels}, \citenamefont {Ki},
  \citenamefont {Watanabe}, \citenamefont {Taniguchi}, \citenamefont
  {Stampfer}, \citenamefont {Guinea},\ and\ \citenamefont
  {Morpurgo}}]{Couto2014Oct}%
  \BibitemOpen
  \bibfield  {author} {\bibinfo {author} {\bibfnamefont {N.~J.~G.}\
  \bibnamefont {Couto}}, \bibinfo {author} {\bibfnamefont {D.}~\bibnamefont
  {Costanzo}}, \bibinfo {author} {\bibfnamefont {S.}~\bibnamefont {Engels}},
  \bibinfo {author} {\bibfnamefont {D.-K.}\ \bibnamefont {Ki}}, \bibinfo
  {author} {\bibfnamefont {K.}~\bibnamefont {Watanabe}}, \bibinfo {author}
  {\bibfnamefont {T.}~\bibnamefont {Taniguchi}}, \bibinfo {author}
  {\bibfnamefont {C.}~\bibnamefont {Stampfer}}, \bibinfo {author}
  {\bibfnamefont {F.}~\bibnamefont {Guinea}},\ and\ \bibinfo {author}
  {\bibfnamefont {A.~F.}\ \bibnamefont {Morpurgo}},\ }\bibfield  {title}
  {\bibinfo {title} {{Random Strain Fluctuations as Dominant Disorder Source
  for High-Quality On-Substrate Graphene Devices}},\ }\href
  {https://doi.org/10.1103/PhysRevX.4.041019} {\bibfield  {journal} {\bibinfo
  {journal} {Phys. Rev. X}\ }\textbf {\bibinfo {volume} {4}},\ \bibinfo {pages}
  {041019} (\bibinfo {year} {2014})}\BibitemShut {NoStop}%
\bibitem [{\citenamefont {Sonntag}\ \emph {et~al.}(2023)\citenamefont
  {Sonntag}, \citenamefont {Watanabe}, \citenamefont {Taniguchi}, \citenamefont
  {Beschoten},\ and\ \citenamefont {Stampfer}}]{Sonntag2023Feb}%
  \BibitemOpen
  \bibfield  {author} {\bibinfo {author} {\bibfnamefont {J.}~\bibnamefont
  {Sonntag}}, \bibinfo {author} {\bibfnamefont {K.}~\bibnamefont {Watanabe}},
  \bibinfo {author} {\bibfnamefont {T.}~\bibnamefont {Taniguchi}}, \bibinfo
  {author} {\bibfnamefont {B.}~\bibnamefont {Beschoten}},\ and\ \bibinfo
  {author} {\bibfnamefont {C.}~\bibnamefont {Stampfer}},\ }\bibfield  {title}
  {\bibinfo {title} {{Charge carrier density dependent Raman spectra of
  graphene encapsulated in hexagonal boron nitride}},\ }\href
  {https://doi.org/10.1103/PhysRevB.107.075420} {\bibfield  {journal} {\bibinfo
   {journal} {Phys. Rev. B}\ }\textbf {\bibinfo {volume} {107}},\ \bibinfo
  {pages} {075420} (\bibinfo {year} {2023})}\BibitemShut {NoStop}%
\bibitem [{\citenamefont {Taychatanapat}\ \emph {et~al.}(2013)\citenamefont
  {Taychatanapat}, \citenamefont {Watanabe}, \citenamefont {Taniguchi},\ and\
  \citenamefont {Jarillo-Herrero}}]{Taychatanapat2013Apr}%
  \BibitemOpen
  \bibfield  {author} {\bibinfo {author} {\bibfnamefont {T.}~\bibnamefont
  {Taychatanapat}}, \bibinfo {author} {\bibfnamefont {K.}~\bibnamefont
  {Watanabe}}, \bibinfo {author} {\bibfnamefont {T.}~\bibnamefont
  {Taniguchi}},\ and\ \bibinfo {author} {\bibfnamefont {P.}~\bibnamefont
  {Jarillo-Herrero}},\ }\bibfield  {title} {\bibinfo {title} {{Electrically
  tunable transverse magnetic focusing in graphene}},\ }\href
  {https://doi.org/10.1038/nphys2549} {\bibfield  {journal} {\bibinfo
  {journal} {Nat. Phys.}\ }\textbf {\bibinfo {volume} {9}},\ \bibinfo {pages}
  {225} (\bibinfo {year} {2013})}\BibitemShut {NoStop}%
\bibitem [{\citenamefont {Albrecht}\ \emph {et~al.}(2017)\citenamefont
  {Albrecht}, \citenamefont {Moers},\ and\ \citenamefont
  {Hermanns}}]{Albrecht2017}%
  \BibitemOpen
  \bibfield  {author} {\bibinfo {author} {\bibfnamefont {W.}~\bibnamefont
  {Albrecht}}, \bibinfo {author} {\bibfnamefont {J.}~\bibnamefont {Moers}},\
  and\ \bibinfo {author} {\bibfnamefont {B.}~\bibnamefont {Hermanns}},\
  }\bibfield  {title} {\bibinfo {title} {{HNF - Helmholtz Nano Facility}},\
  }\href {https://doi.org/10.17815/jlsrf-3-158} {\bibfield  {journal} {\bibinfo
   {journal} {Journal of large-scale research facilities JLSRF}\ }\textbf
  {\bibinfo {volume} {3}},\ \bibinfo {pages} {112} (\bibinfo {year}
  {2017})}\BibitemShut {NoStop}%
\end{thebibliography}
\end{document}